\documentclass[largeformat,article,a4paper]{interact}

\usepackage{epstopdf}
\usepackage{subfigure}

\usepackage[numbers,sort&compress]{natbib}
\bibpunct[, ]{[}{]}{,}{n}{,}{,}
\makeatletter
\def\NAT@def@citea{\def\@citea{\NAT@separator}}
\makeatother

\theoremstyle{plain}

\theoremstyle{definition}

\theoremstyle{remark}

\usepackage{color,soul}

\def\bx{\boldsymbol{x}}
\def\bu{\boldsymbol{u}}
\def\br{\boldsymbol{r}}
\def\bk{\boldsymbol{k}}
\def\be{\boldsymbol{e}}
\def\bn{\boldsymbol{n}}
\def\bN{\boldsymbol{N}}

\def\bomega{\boldsymbol{\omega}}
\def\bOmega{\boldsymbol{\Omega}}

\DeclareMathAlphabet{\mathbfsf}{\encodingdefault}{\sfdefault}{bx}{n}

\begin{document}

\articletype{}

\title{Small-scale anisotropy induced by spectral forcing and by rotation in non-helical and helical turbulence}

\author{
\name{D. Vallefuoco\textsuperscript{a}, A. Naso\textsuperscript{a} and F.~S. Godeferd\textsuperscript{a}}
\affil{\textsuperscript{a}Laboratoire de M\'ecanique des Fluides et d'Acoustique, CNRS, \'Ecole Centrale de Lyon, Universit\'e Claude Bernard Lyon 1 and INSA de Lyon, 36 avenue Guy de Collongue, F-69134 \'Ecully Cedex, France}
}

\maketitle

\begin{abstract}
We study  the effect of large-scale spectral forcing on the scale-dependent anisotropy of the velocity field
in direct numerical simulations of homogeneous incompressible turbulence. Two forcing methods are considered: the steady ABC single wavenumber scheme and the unsteady non-helical or helical Euler scheme. The results are also compared with high resolution data obtained with the negative viscosity scheme.
A fine-grained characterization of anisotropy, consisting in measuring some quantities related to the two-point velocity correlations, is used: we perform a modal decomposition of the spectral velocity tensor into energy, helicity and polarization spectra. Moreover, we include the explicit dependence of these three spectra on the wavevector direction.  
The conditions that allow anisotropy to develop in the small scales due to forcing alone are clearly identified. It is shown that, in turbulent flows expected to be isotropic, the ABC forcing yields  significant energy and helicity directional anisotropy down to the smallest resolved scales, like the helical Euler scheme when an unfavourable forcing scale is used.
The direction- and scale-dependent anisotropy is then studied in rotating turbulence.
It is first shown that, in the ABC-forced simulations the slope of the energy spectrum is altered and the level of anisotropy is similar to that obtained at lower Rossby number in Euler-forced runs, a result due both to the nature of the forcing itself and to the fact that it allows an inverse cascade to develop. 
Second, we show that, even at low rotation rate, the natural anisotropy induced by the Coriolis force is visible at all scales.
Finally, we identify two different wavenumber ranges in which anisotropy behaves differently, and show that the characteristic lenghscale separating them is not the Zeman scale. 
If the Rossby number is not too low, this scale is the one at which rotation and dissipation effects balance.
\end{abstract}

\section{Introduction}\label{sec:intro}

According to the classical Kolmogorov K41 theory \cite{bib:Kolmogorov}, 
 in turbulent flows at asymptotically large Reynolds number, the large-scale dynamics should affect small scales statistical properties only through the energy production rate, \textit{i.e.} small scales should be statistically independent of large scales, and have a universal behaviour.
This assumption, referred to as the local isotropy hypothesis, has been studied by many authors but 
is still controversial.
These authors agree about energy cascading from large to small scales mainly through local triadic interactions in Fourier space.
However, some works also showed that the energy-containing scales directly affect the small scales dynamics through distant triadic interactions, \textit{i.e.} nonlinear interactions corresponding to wavenumber triangles with very large scale separation.
In particular, \cite{bib:Brasseur1987,1991tsf.....1...16B} considered the nonlinear term in Fourier space and analysed the nonlinear interactions among modes in a single triad with a wavenumber in the energy-containing scales. It was shown that the contribution related to such a triad does not vanish in the asymptotic limit of infinite scale separation, and thus it was argued that small scales are not independent of large scales in the asymptotic limit of large Reynolds number.
Yeung \& Brasseur (1991) \cite{bib:yeung1991} confirmed this analysis by observing small scale anisotropy in numerical simulations with strongly anisotropic large scale forcing. In fact, since small scale anisotropy was found to increase with the wavenumber and to be consistent with the distant triad equations, local anisotropy should therefore persist at asymptotically high Reynolds number.
The local isotropy hypothesis was also shown to be violated in homogeneous sheared turbulence by the measurement of statistical quantities in the physical space in direct numerical simulations (DNS) \cite{bib:PumirShraiman,bib:Pumir96} and in experiments \cite{bib:ShenWarhaft}.
The detailed structure of small scales in highly anisotropically forced turbulence was also investigated by \cite{bib:yeung1995} in both Fourier and physical space. Anisotropic redistribution of energy and phase in high wavenumber shells was predicted and observed in DNS. In particular, a reduction of energy was detected in the directions of the forcing wavenumbers.
In any case, the study of anisotropic turbulence and of its scale-dependent features through classical Fourier analysis requires to disentangle the effect of physical sources of anisotropy from those of other artificial mechanisms, like energy and helicity production in forced simulations. Identifying and quantifying the anisotropy induced by some widely used forcing schemes in turbulence intended to be isotropic is the first concern of this work.
Our second concern is to investigate homogeneous non-helical and helical forced turbulence subject to a background rotation, by characterizing its scale- and angle-dependent anisotropy.
The motivation comes for instance from previous studies of freely decaying rotating turbulence \cite{bib:delache2014}, in which a refined anisotropic characterization was absolutely required to understand the subtle effect of the Coriolis force on each scale of the flow.
However, the drop in Reynolds number was severe due to dissipation, so that forced rotating turbulence should rather be considered.

In order to study statistically stationary turbulence, many velocity forcing schemes have been used so far in numerical simulations. In particular, large-scale spectral forcing methods were used in homogeneous spectral simulations and consist in providing energy to the low wavenumber modes, which is consistent with the concept of Richardson cascade, see \textit{e.g.} \cite{bib:eswaran1988}, \cite{bib:sullivan1994}, \cite{bib:alvelius1999}.
However, since only a finite number of wavenumbers is excited in these simulations, anisotropy may develop at large scales and eventually branch out to smaller scales.
Detecting this kind of anisotropy requires direction-dependent statistics.
In this paper we analyse three forcing schemes representative of large-scale forcing, the Euler, the ABC and the negative viscosity forcing methods  \cite{bib:pumir1994,bib:mininni2009,bib:jimenez1993}.
The impact of these  forcing schemes on the produced turbulence scale-dependent anisotropy is characterized by
a modal decomposition of the spectral velocity tensor, such that it depends only on energy, helicity and polarization  spectral densities, which in turn depend on the orientation as well as on the modulus of the wavevector.
Note that while in \cite{bib:yeung1991,bib:yeung1995} an explicitly and highly anisotropic forcing was used, we investigate here the unwanted intrinsic anisotropy of large scale spectral forcing schemes.

After assessing this in turbulence intended to be isotropic, we choose to extend our study to the context of rotating homogeneous turbulence. This context is relevant for instance to geophysical and industrial flows, or academic configurations such as the von K\'arm\'an-forced turbulence \cite{bib:vonkarman}. 
It is nowadays commonly admitted that background rotation introduces significant anisotropy in the turbulent dynamics through both linear and nonlinear mechanisms (see \textit{e.g.} \cite{bib:godeferd2015}). 
The flow regime can be characterized by two independent non-dimensional parameters. One possible choice is the Reynolds number $\textit{Re}^L=UL/\nu$ and the macro-Rossby number $\textit{Ro}^L=U/(2\Omega L)$, where $U$ is a large-scale characteristic velocity [\textit{e.g.} the root-mean-square (r.m.s.) velocity], and $L$ is a large-scale characteristic lengthscale (\textit{e.g.} the integral scale), $\nu$ is the kinematic viscosity, and $\Omega$ is the rotation rate.
The micro-Rossby number is defined as $\textit{Ro}^\omega={\omega'}/{(2\Omega)}$, where $\omega'$ is the r.m.s. vorticity.
The macro- and the micro-Rossby numbers quantify the relative importance of the advection with respect to the rotation rate. 
In addition to $L$, three more characteristic lengthscales can be defined: (i) the Kolmogorov scale $\eta=(\nu^3/\epsilon)^{1/4}$, where $\epsilon$ is the mean energy dissipation rate; (ii) the scale at which the inertial timescale $(r^2/\epsilon)^{1/3}$ equals the rotation timescale $1/\Omega$, $r_\Omega=\sqrt{\epsilon/(2\Omega)^3}$ \cite{bib:zeman, bib:woods_1974}, which is referred to as the Zeman scale; (iii) the scale at which the dissipative timescale $r^2/\nu$ equals the rotation timescale, $r_{\Omega d}=\sqrt{\nu/(2\Omega)}$. From the above definitions of $\eta$ and $r_\Omega$, $r_{\Omega d}=r_\Omega^{1/3}\eta^{2/3}$.
One alternative choice for the independent parameters may be two characteristic lengthscale ratios. 
Furthermore, by setting $\epsilon\sim U^3/L$, the ratio of the integral scale to the Kolmogorov scale and the ratio of Zeman scale to the integral scale are linked to $\textit{Re}^L$ and $\textit{Ro}^L$: $L/\eta\sim{\textit{Re}^L}^{3/4}$ and $r_\Omega/L\sim{\textit{Ro}^L}^{3/2}$.
Similarly, if $\omega'\sim \nu/\eta^2$, $\textit{Ro}^\omega\sim(r_\Omega/\eta)^{2/3}$ and $\epsilon\sim U^3/L$ also leads to $\textit{Ro}^\omega\sim({\textit{Re}^L})^{1/2}\textit{Ro}^L$.
The assumption that $\epsilon\sim U^3/L$ at high Reynolds numbers has been extensively investigated in isotropic turbulence and a precise scaling law for $C_\epsilon=\epsilon/(U^3 L)$ has been obtained for non-equilibrium (\textit{e.g.} decaying) turbulence (see \cite{Vassilicos2015} for a review), but for forced turbulence $C_\epsilon$ has been found to be constant and independent of the forcing scheme and the forcing wavenumber, even if turbulence is quasi-periodic and time averages are considered, see \cite{bos2007spectral,Goto20151144}.
Note that $\textit{Ro}^\omega$ (or the equivalent parameters $r_\Omega/\eta$ and $\textit{Re}^L{\textit{Ro}^L}^2$) does not depend on large-scales characteristic quantities like the integral lengthscale or the r.m.s. velocity, and is indeed the only nondimensional parameter that arises from a dimensional analysis if only $\epsilon$, $\nu$ and $\Omega$ are taken into account.
If $\nu$ tends to zero (and the Reynolds number tends to infinity), both $\eta$ and $r_{\Omega d}$ tend to zero. The only relevant small-scale characteristic lengthscale is then $r_\Omega$.
For this reason, according to classical dimensional arguments \cite{bib:mininni2012,bib:delache2014,bib:zeman}, in the asymptotically inviscid limit scales much larger than $r_\Omega$ are mainly affected by rotation while scales much smaller than $r_\Omega$ are dominated by the nonlinear dynamics and are expected to recover isotropy.
In the following sections we will refer to characteristic wavenumbers instead of length scales: $k_\eta=1/\eta$, $k_\Omega=1/r_\Omega$ and $k_{\Omega d}=1/r_{\Omega d}$.
We characterize the anisotropy that naturally arises because of rotation through the same scale- and angle-dependent statistics we use to detect artificial anisotropy of forcing schemes in ``isotropic'' turbulence. Different flow regimes in terms of Rossby and Reynolds numbers, as well as the possibility of helicity injection are considered.

The paper is organized as follows. In section \ref{sec:forc} we introduce the refined two-point statistics used in the following as diagnostics for anisotropy characterization and present the numerical simulation method, the Euler and ABC  forcing schemes.
In section \ref{sec:results1} we compare the two forcing methods and identify the conditions that allow anisotropy to develop at small scales in the non-rotating case. 
In order to show the effect of an increase in Reynolds number we also study data from $2048^3$ resolution DNS forced through the negative viscosity method \cite{bib:kanedajot}. Section \ref{sec:results2} is devoted to the characterization of anisotropy induced by background rotation in homogeneous non-helical and helical turbulence.
In section \ref{sec:kT} two different anisotropic ranges are identified, and a physical interpretation of the separating scale is provided.
Conclusions are drawn in section~\ref{sec:concl}.

\section{Methodology}

\label{sec:forc}
We consider an incompressible fluid whose motion follows the Navier-Stokes equations
\begin{align}
\begin{split}
\frac{\partial \boldsymbol{u}}{\partial t}+(\boldsymbol{\omega}+2\boldsymbol{\Omega})\times \boldsymbol{u}=-\nabla P+\nu \nabla^2\boldsymbol{u}+\boldsymbol{F}  \label{eq:NS}\\
\nabla \cdot \boldsymbol{u}=0
\end{split}
\end{align}
where $\boldsymbol{u}$ is the velocity field, $\boldsymbol{\omega}=\nabla \times \boldsymbol{u}$ is the vorticity, $P$ is the total pressure (sum of the hydrodynamic pressure and of the centrifugal contribution) divided by density, $\nu$ is the kinematic viscosity, $\boldsymbol{F}$ is an external force, $\boldsymbol{\Omega}$ is the possible rotation rate of the frame, and $-2\boldsymbol{\Omega}\times \boldsymbol{u}$ is therefore the Coriolis force.

In the present section we first describe in detail the statistical indicators that will be used thereafter to evaluate the scale- and direction-dependent anisotropy of the velocity field. We will then describe the numerical set-up and the forcing schemes.

\subsection{Fine-grained anisotropy in two-point statistics}
\label{sec:twopointstats}
The characterization of anisotropy in homogeneous turbulence addresses a two-fold question. First, what physical quantities are suitable to qualitatively detect isotropy breaking in turbulence subject to external distorsions such as solid body rotation, density gradient, mean shear, \textit{etc.}? Second, how does one quantify and compare the level of anisotropy? One therefore needs a relevant characterization of this anisotropy, and several choices are possible. 

Considering the Reynolds stress tensor $\mathbfsf{{R}}$ of components $R_{ij}(\br,t)=\langle u_i(\bx,t) u_j(\bx+\br,t)\rangle$, where $\bx=\left( x_1,x_2,x_3 \right)$ is the Cartesian coordinate in physical space, $\boldsymbol{r}$ is the separation vector, $t$ is time and $\langle\,\rangle$ represents ensemble averaging, one can obtain the components of the anisotropic part of $\mathbfsf{{R}}$, $b_{ij}=R_{ij}/R_{kk}-\delta_{ij}/3$ (the Einstein summation convention is used here). If the off-diagonal components of $\mathbfsf{b}$ are not zero the flow is anisotropic, but these quantities only represent anisotropy from a global point of view---mostly related to the large scales. A widely adopted characterization of anisotropy based on $\mathbfsf{b}$ is the method proposed by Lumley \& Newman (1977) \cite{bib:lumley77} which consists in identifying the dominant structure of the flow from the position of the second and third invariants ($I_2$,$I_3$) of $\mathbfsf{b}$ within the so-called Lumley triangle. This tells if the flow structure is mostly 2-component axisymmetric, 1-component, or isotropic, depending on the closeness of the  ($I_2$,$I_3$) point  to one of the vertices of the triangle. However, useful as this simple method may be, it does not tell which scales are most anisotropic.  A refined picture is for instance required for  rotating turbulence in which one has to identify isotropic and anisotropic subranges at different length scales (see section~\ref{sec:results2}) \cite{bib:zeman, bib:lamriben2011, bib:mininni2012, bib:delache2014}.

We therefore introduce hereafter a scale-by-scale evaluation of anisotropy. In addition to the lengthscale or wavenumber, we also retain the dependence of the spectra on the polar angle about the axis of symmetry. This description is suitable for a wide range of statistically axisymmetric flows, such as turbulence subject to solid body rotation, stratified turbulence, flows subject to axisymmetric contractions or expansions or more generally axisymmetric strain, magneto-hydrodynamic turbulence for a conducting fluid subject to an external magnetic field of fixed orientation. Non axisymmetric cases are more complex and only a few studies have been devoted to their statistical description.

\subsubsection{Modal decomposition of the Reynolds-stress tensor spectrum}\label{sec:decompR}
\label{sec:modal}
\begin{figure}[]
        \centering
                \includegraphics[width=.3\textwidth]{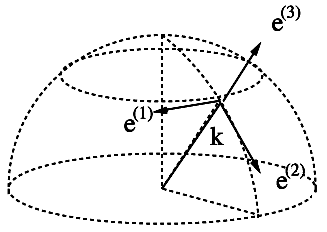}
        \caption{\label{fig:Craya}  Craya-Herring frame of reference.} 
\end{figure}

Since we deal with homogeneous turbulent flows, the two-point correlation tensor  $\mathbfsf{{R}}$  is independent of $\bx$, and---if it tends to zero sufficiently rapidly as $|\br|$ increases---we can consider its Fourier transform 
\begin{equation}
\hat{R}_{ij}(\bk)=\frac{1}{(2\pi)^{3}}\iiint R_{ij}(\br)\mathrm{e}^{-\mathrm{i}\bk\cdot\br}\mathrm{d}^3\br
\label{eq:defRhat}
\end{equation}
(for simplicity, we drop here the dependence upon time $t$).
Note that the incompressibility condition $\nabla \cdot \boldsymbol{u}=0$ implies $\partial R_{ij}(\br)/\partial r_j =0$, which by Eq. \eqref{eq:defRhat} leads to $\hat{R}_{ij}(\bk)k_j=0$. Furthermore, since $R_{ij}(\br)$ is real and $R_{ij}(\br)=R_{ji}(-\br)$ from its definition, $\hat{R}_{ij}(\bk)$ is a Hermitian matrix, \textit{i.e.} $\hat{R}_{ij}^*(\bk)=\hat{R}_{ji}(\bk)$, where $^*$ stands for complex conjugate.
It is useful to project the tensor $\hat{\mathbfsf{R}}$ onto a polar-spherical orthonormal basis ($\be^{(1)}$,$\be^{(2)}$,$\be^{(3)}$) defined from the vector $\bn$ bearing the axis of symmetry, with 
\begin{equation}
\boldsymbol{e}^{(1)}=\frac{\boldsymbol{k} \times \boldsymbol{n}}{|\boldsymbol{k} \times \boldsymbol{n}|}
,\quad
\boldsymbol{e}^{(2)}=\boldsymbol{e}^{(3)} \times \boldsymbol{e}^{(1)}
,\quad
\boldsymbol{e}^{(3)}=\frac{\boldsymbol{k}}{k},
\end{equation}
which is the so-called Craya-Herring frame \cite{craya}, see Fig. \ref{fig:Craya}. $\boldsymbol{e}^{(1)}$ and $\boldsymbol{e}^{(2)}$ are respectively referred to as \emph{toroidal} and \emph{poloidal} directions. By enforcing incompressibility and Hermitian symmetry,
\begin{equation}
\hat{R}_{ij}(\bk)=\Phi^{1}(\boldsymbol{k})\boldsymbol{e}^{(1)}\boldsymbol{e}^{(1)}+ \Phi^{12}(\boldsymbol{k}) \boldsymbol{e}^{(1)}\boldsymbol{e}^{(2)}+ \Phi^{12*}(\boldsymbol{k}) \boldsymbol{e}^{(2)}\boldsymbol{e}^{(1)} + \Phi^{2}(\boldsymbol{k}) \boldsymbol{e}^{(2)}\boldsymbol{e}^{(2)},
\label{eq:craya_decompR}
\end{equation}
where $\Phi^{1}/2$ and $\Phi^{2}/2$ are the toroidal and the poloidal energy spectral densities, respectively.
Equation \eqref{eq:craya_decompR} can be rewritten as \cite{bib:sagaut2008, bib:CMG}
\begin{equation}
\hat{\mathsf{R}}_{ij}(\bk)=e(\bk)P_{ij}(\bk)+\Re\left( {z}(\bk)N_i(\bk)N_j(\bk)\right)+\mathrm{i}h(\bk)\epsilon_{ijl}\frac{k_l}{2k^2}, \label{eq:decomp}
\end{equation}
where $P_{ij}=\delta_{ij}-k_ik_j/k^2$ is the projector onto the ($\be^{(1)}$,$\be^{(2)}$) plane, $\bN(\bk)=\be^{(2)}(\bk)-\mathrm{i}\be^{(1)}(\bk)$ are helical modes \cite{bib:waleffe}, $\epsilon_{ijk}$ is the alternating Levi-Civita tensor and $\Re$ denotes the real part.
The decomposition (\ref{eq:decomp}) displays three important spectral functions which characterize fully the second-order velocity correlations of the flow and carry useful physical meaning about the flow structure at different scales \cite{bib:CMG, bib:sagaut2008, bib:delache2014}:

\begin{enumerate}
\item $e(\bk)=\hat{\mathsf{R}}_{ii}(\bk)/2=\left(\Phi^1(\bk)+\Phi^2(\bk)\right)/2$ is the spectral energy density, and upon integration over spherical shells of radius $k=|\bk|$ provides the kinetic energy spectrum $E(k)=\int \! e(\boldsymbol{k}) \delta (|\boldsymbol{k}|-k) \,  \mathrm{d}\boldsymbol{k}$, that scales as $k^{-5/3}$ in the inertial range of high Reynolds number isotropic turbulence according to the Kolmogorov theory. 
If energy is concentrated in modes corresponding to wavevectors close to the plane $\bk \cdot \bn=0$, the flow is almost bidimensional,
while energy concentrated in wavevectors close to $\bn$ indicates a trend towards a vertically-sheared horizontal flow.

\item The complex-valued function 
\begin{equation}
{z}(\bk)=\left(  \Phi^2(\bk)-\Phi^1(\bk)\right)/2 + \mathrm{i} \Re \Phi^{12}(\bk) \label{eq:Z} 
\end{equation}
is the polarization spectral density and contains information on the structure of the flow at different scales. Consider for instance a shell of radius $k$ in spectral space in which the wavevectors closer to the horizontal plane $k_z=0$ hold much more energy than the others (which is the case of strongly rotating turbulence if $\bOmega$ is parallel to the $x_3$ axis). In this special case, if the real part of polarization is mostly dominated by the poloidal spectral energy $\Phi^2$, the corresponding flow structure at the scale $1/k$ is characterized by axial velocity, or ``jetal'' structures, whereas if $\Phi^1$ prevails, axial vorticity is more important and the flow displays ``vortical'' structures.
Detailed comments about the role of ${z}$ in rotating turbulence or MHD turbulence can be found in \cite{bib:CMG, bib:favier,bib:delache2014}.

In section \ref{sec:results1} we show normalised integrated spectra of the real part  of $z(\boldsymbol{k})$,
$\Re{Z}(k)/E(k)=\left(E^\text{pol}(k)-E^\text{tor}(k)\right)/E(k)$, 
where 
$E^\text{pol}(k)=\int \! \Phi^2(\bk) \delta (|\boldsymbol{k}|-k) \, \mathrm{d}\boldsymbol{k}$ 
and 
$E^\text{tor}(k)=\int \! \Phi^1(\bk)  \delta (|\boldsymbol{k}|-k) \, \mathrm{d}\boldsymbol{k}$. 

\item Finally, $h(\bk)=2k \Im \Phi^{12}(\bk)$, where $\Im$ stands for the imaginary part, is the helicity spectral density. 
In physical space, helicity density is the scalar product between velocity and vorticity, $\boldsymbol{u} \cdot \boldsymbol{\omega}$, and---exactly like energy---its integral is an inviscid invariant \cite{bib:moffatt1969,bib:moffatt1992} (even in the presence of background rotation).
$h(\bk)$ is the Fourier transform of the velocity-vorticity correlation $\left<\bu(\bx)\cdot\bomega(\bx+\br)\right>$, and thus $\int \! h(\bk) \, \mathrm{d}\boldsymbol{k}$ equals the mean helicity. 
The helicity spectrum is
\begin{equation}
\label{eq:H}
H(k)=\int \! h(\bk) \delta (|\boldsymbol{k}|-k) \, \mathrm{d}\boldsymbol{k}.
\end{equation}
Since helicity is a pseudoscalar quantity, any turbulent flow with non-vanishing mean helicity lacks mirror-symmetry. However, in sections \ref{sec:results1} and \ref{sec:results2} we will focus on directional and polarization anisotropy, and the word ``anisotropic'' will refer to any isotropy breaking but mirror-symmetry breaking.
\end{enumerate}

\subsubsection{Directional dependence of the spectra}\label{sec:dirspectra}
%
In the above decomposition we have retained the general $\bk$ dependence.  Furthermore, one can use axisymmetry to consider only the dependence of the spectra upon the axial and horizontal components of the wavevector $\bk$ (see for instance \cite{bib:galtier}), or upon the wavenumber $k$ and the polar orientation $\theta$ of $\bk$ with respect to the axis of symmetry \cite{bib:cambon1989,bib:godeferd2003}.
Therefore, in our following analysis of spectral anisotropy, we shall present  $\theta$-dependent spectra,  discretizing $k$ between minimal and maximal values set by the computational box size and the resolution, and considering angular averages of spectra in five angular sectors in the interval $\theta \in [0,\pi/2]$, \textit{i.e.} $[(i-1)\pi/10,i\pi/10]$ with $i=1,\cdots,5$.
We call $E_i(k)$, $H_i(k)$ and $\Re Z_i(k)$ the spectra of energy, helicity and real part of polarization. They are obtained by partial integration of the corresponding spectral densities over these sectors.
Note that the spectra for all angular sectors are normalised such that for directionally isotropic turbulence they collapse onto the corresponding spherically-integrated spectrum, \textit{e.g.} the $E_i(k)$ spectra collapse on $E(k)$. 
The limited number of sectors is imposed by the need of a minimal number of discrete wavevectors in every sector for achieving  decent sample size from DNS data. Even so, in the small wavenumbers, very few wavevectors lie within the averaging regions, but this is a known fact for all direct numerical simulations based on pseudo-spectral schemes.

Finally, note that the directional spectra $E_i$, $H_i$, $\Re Z_i$ carry the most accurate scale-by-scale information about the statistically axisymmetric flow second-order statistics, but other choices could be made.
We only recall here the fact that the anisotropy tensor $\mathbfsf{b}$, which carries a rough information on anisotropy, can be split as $b_{ij}=b_{ij}^{(e)}+b_{ij}^{(z)}+b_{ij}^{(h)}$ into more informative contributions brought up by integrating the spectra:
\begin{align*}
b_{ij}^{(e)}&=\frac{1}{\left<u_ku_k\right>}\iiint\left[e(\bk)-E(k)/(4\pi k^2)\right]P_{ij}d^3\bk  \\
b_{ij}^{(z)}&=\frac{1}{\left<u_ku_k\right>}\iiint \Re\left[z(\bk)N_i(\bk)N_j(\bk)\right]d^3\bk  \\
b_{ij}^{(h)}&=\frac{1}{\left<u_ku_k\right>}\iiint\mathrm{i}h(\bk)\epsilon_{ijl}\frac{k_l}{2k^2}d^3\bk \\
&=\frac{1}{\left<u_ku_k\right>}\iiint\mathrm{i}\frac{H(k)}{4\pi k^2}\epsilon_{ijl}\frac{k_l}{2k^2}d^3\bk +\frac{1}{\left< u_ku_k\right>}\iiint\mathrm{i}\left[h(\bk)-\frac{H(k)}{4\pi k^2}\right]\epsilon_{ijl}\frac{k_l}{2k^2}d^3\bk.
\end{align*}
For instance, in exactly isotropic mirror-symmetric three-dimensional turbulence, $b_{ij}=b_{ij}^{(e)}=b_{ij}^{(z)}=b_{ij}^{(h)}=0$,
whereas  two-dimensional turbulence (for which helicity is identically zero) with only two components of velocity in the plane ($1$,$2$) is characterized as the departure from 3D isotropy by $b_{33}=-1/3$, $b_{33}^{(e)}=1/6$ and $b_{33}^{(z)}=-1/2$ \cite{bib:CMG}. Thus the $e$, $z$, $h$-related contributions to the deviatoric tensor $\mathbfsf{b}$ provide useful quantitative indicators about anisotropic trends in the flow, but retaining the spectral information permits to qualify the flow structure in a scale-dependent way.

\subsection{Numerical set-up and forcing schemes}
The Navier-Stokes equations \eqref{eq:NS} are solved in a three-dimensional $2\pi$--periodic cube $C$ with a classical Fourier pseudo-spectral algorithm (see for instance \cite{bib:orszag,bib:vincent}). The code uses the 2/3-rule for dealiasing and  third-order Adams-Bashforth scheme for time marching.

The periodic velocity field $\boldsymbol{u}(\boldsymbol{x})$ can be expanded as an infinite Fourier series
\begin{equation}
\bu(\bx)=\sum\limits_{\bk}\hat{\bu}(\bk)e^{\mathrm{i} \bk\cdot\bx}
\end{equation}
where $\bk$ represents now discrete wavevectors and $\hat{\bu}(\bk)=(2\pi)^{-3}\int_C\bu(\bx)e^{-\mathrm{i} \bk\cdot\bx}\mathrm{d}\boldsymbol{x}$ are the Fourier coefficients of $\bu(\bx)$.
$\hat{\bu}(\bk)$ can be projected onto the Craya-Herring frame,
\begin{equation}\label{eq:craya_decomp}
\boldsymbol{\hat{u}}(\boldsymbol{k})=u^{(1)}(\boldsymbol{k})\boldsymbol{e}^{(1)}(\boldsymbol{k})+u^{(2)}(\boldsymbol{k})\boldsymbol{e}^{(2)}(\boldsymbol{k})
\end{equation}
with no component of $\hat{\bu}$ along $\be^{(3)}$ because of the incompressibility condition $\bk\cdot\hat{\bu}(\bk)=0$.
$R_{ij}(\br)$ is periodic too, and the tensor $E_{ij}(\bk)=\langle\hat{u}_i(\bk)\hat{u}_j^*(\bk)\rangle$ represents its Fourier coefficients. The decomposition developed in section \ref{sec:decompR} for $\hat{R}_{ij}$ may be repeated for $E_{ij}$ with no formal difference.
In addition, the spectral densities appearing in Eq. \eqref{eq:decomp} are now linked to $\hat{\bu}(\bk)$, \textit{i.e.}
$e(\bk)=\langle\hat{\bu}(\bk)\cdot\hat{\bu}^*(\bk)\rangle/2$, $h(\bk)=\langle\hat{\bu}(\bk)\cdot\hat{\bomega}^*(\bk)\rangle$, ${z}(\bk)=\langle u^{(2)}(\bk)u^{(2)*}(\bk)-u^{(1)}(\bk)u^{(1)*}(\bk)\rangle/2 + \mathrm{i}\langle u^{(1)}_R(\bk)u^{(2)}_R(\bk) + u^{(1)}_I(\bk)u^{(2)}_I(\bk) \rangle$, where the subscripts $_R$ and $_I$ stand for real and imaginary parts.
We compute the spherically integrated spectra as sums of the corresponding spectral densities in unitary-thickness shells.
From the definition of vorticity and the Schwarz inequality one can show that a realizability condition holds: $|h(\boldsymbol{k})|\le 2\,|\boldsymbol{k}| \, e(\boldsymbol{k})$.
Therefore, we define \textit{relative} helicity as ${H}_\text{rel}={\left<{H}\right>L_h}/{K}$
where $K=\sum  e(\boldsymbol{k})$ is the turbulent kinetic energy and $L_h$ is a modified lengthscale (different from the integral lengthscale), defined from the spherically integrated kinetic energy spectrum as
\begin{equation}
L_h=\frac{1}{2}\frac{\sum E(k_i) }{\sum k_iE(k_i) }
\end{equation}
so that, from the above inequality, ${H}_\text{rel}\le 1$.

When performing direct numerical simulations, one would like  to force turbulence for two reasons. First, it permits to reach higher Reynolds numbers than in freely decaying turbulence.
Second, under some assumptions, statistics can be obtained with time-averaging rather than ensemble averaging (see \textit{e.g.} \cite{bib:mathieu_scott_2000}) which would be very costly considering the fact that our refined statistics require a large number of samples. 
Therefore, the velocity field obtained by DNS is processed to obtain the required statistics, presented in section~\ref{sec:twopointstats}, and these statistics are time-averaged over a few eddy-turnover-times of turbulence after the initial transient is passed, when turbulence has reached a statistically steady state.

The goal of forcing turbulence is to represent, as a model force, the essential features of forcing mechanisms in more complex turbulent flows, and to reproduce, without simulating complete complex systems, situations of actual flows, such as \textit{e.g.} injection of energy by large-scale instabilities in atmospheric flows, or stirring devices in industrial flows. For instance, the well-known von K\'arm\'an experiment consists of two counter rotating rotors \cite{bib:vonkarman}, that not only inject energy at large scales in the flow, but also helicity. 
For this reason, we wish to investigate the possibility of representing these mechanisms through simple models and to study their impact on the anisotropy of the flow, including the possibility of injection of helicity.
 We choose in this work to consider the ABC forcing (see for instance \cite{bib:mininni2012} in hydrodynamic turbulence or \cite{bib:brummell} in magnetohydrodynamic turbulence), and the Euler forcing \cite{bib:pumir1994}. 
 
Isotropic turbulence data obtained by a negative viscosity forcing scheme in Kaneda's group \cite{bib:kanedajot} will also be considered hereafter.
According to this method, the term ${\bf F}$ of Eq. (\ref{eq:NS}) has the same form as the dissipative term, but its ``viscosity coefficient'' is negative for all the modes with wavenumbers $|\boldsymbol{k}|\le k_F$, and zero for the modes such that $|\boldsymbol{k}|> k_F$ ($k_F$ is the wavenumber separating forced and unforced spectral ranges).
We now describe the two other forcing schemes used in the present investigation.

\subsubsection{ABC forcing}
The ABC forcing  consists in adding in the Navier-Stokes equations (\ref{eq:NS}) an external force $\boldsymbol{F}_\text{ABC}$ corresponding to an Arnold-Beltrami-Childress flow (see \textit{e.g.} \cite{bib:childress1995}):
\begin{multline}\label{FABC}
\boldsymbol{F}_\text{ABC}=[B \cos(k_F\, y)+C\sin(k_F\, z)]\, \boldsymbol{\hat{\imath}}+[C\cos(k_F\, z)+A\sin(k_F\, x)]\, \boldsymbol{\hat{\jmath}} \\+[A\cos(k_F\, x)+B\sin(k_F\, y)]\, \boldsymbol{\hat{k}},
\end{multline}
for a given large scale wavenumber $k_F$.
Since $\boldsymbol{F}_\text{ABC}$ is an eigenfunction of the curl operator with eigenvalue $k_F$, the ABC forcing injects helicity, in addition to energy, in the flow.
For the ABC-forced runs reported in sections \ref{sec:results1} and \ref{sec:results2}, \mbox{$A=B=C$}.
For the sake of simplicity let the constants $A$, $B$ and $C$ be equal to $2$. Then in Fourier space the expression \eqref{FABC} becomes
\begin{flalign}
\begin{split}
& \begin{array}{cllcrcllcrccc} \boldsymbol{\hat{F}}_\text{ABC} & = & [0 & \pm \mathrm{i}  &1]& \mbox{if } \boldsymbol{k}&=&[\mp k_F & 0& 0]&  \\  \boldsymbol{\hat{F}}_\text{ABC} & = & [1 & 0 &\pm \mathrm{i}]& \mbox{if } \boldsymbol{k}&=&[0 & \mp k_F& 0]&                                               \\  \boldsymbol{\hat{F}}_\text{ABC} & = & [\pm \mathrm{i} & 1 &0]& \mbox{if } \boldsymbol{k}&=&[0 & 0& \mp k_F]&              \end{array}\\ 
& \; \: \boldsymbol{\hat{F}}_\text{ABC}  \;\;\,= \;\;\: [0 \qquad 0 \qquad 0] \qquad \mbox{otherwise.}
\end{split} \label{eq:FABC2}
\end{flalign}
%
In terms of flow structure, the large-scale flow induced by the ABC forcing is very much like Taylor-Green vortices, but extended to three dimensions. More precisely, $\boldsymbol{F}_\text{ABC}$ induces permanent large-scale curved helical rotors associated with a single wavelength. 

Thus, $\boldsymbol{\hat{F}}_\text{ABC}$ is a steady force that excites only six modes and injects a given amount of helicity. In rapidly rotating turbulence, an inverse energy cascade can arise so that it can be difficult to reach a statistically stationary state \cite{bib:mininnicascade}.
Indeed, similarly to ABC-forced rotating simulations present in the literature, for the ABC-forced rotating run analysed in section \ref{sec:results2},  spectra are computed by time-averaging, although a fully statistically stationary state is not reached.

\subsubsection{Non-helical and helical Euler forcing}\label{sec:Euler}

In order to overcome some of the limitations of ABC forcing, we use the Euler forcing, which can be thought of as introducing three-dimensional large-scale vortices that evolve in time by interacting with each other---but not with the other scales of the flow---in a manner closer to actual inviscid turbulent nonlinear dynamics. Unlike the ABC forcing, the external force induced by the Euler scheme is unsteady and chaotic, the number of excited modes depends on $k_F$, and the amount of  injected helicity  can be controlled.

We now describe in detail how the Euler forcing is implemented.
The Euler-forced simulations \cite{bib:pumir1994} are inspired by the truncated Euler dynamics \cite{bib:cichowlas2005}:  the lowest-wavenumbers modes, corresponding to wavevectors $\bk$ such that $0\leq |\boldsymbol{k}|\leq k_F$ ($k_F$ is the largest forcing wavenumber), obey the three-dimensional incompressible Euler equations (possibly with background rotation) and are independent of the other modes.
Of course the modes corresponding to wavenumbers $\bk$ such that $|\boldsymbol{k}|> k_F$ are solutions of the incompressible Navier-Stokes equations and also depend on the modes in the Euler forcing sphere.
For the spherically truncated inviscid system, the quadratic nonlinear term is computed through a convolution in Fourier space so that no aliasing error arises.
Since energy and helicity are conserved within every nonlinear triadic interaction \cite{bib:kraichnan1973}, in this truncated system total energy and helicity are conserved as well.
Background rotation does not affect this conservation property, since the Coriolis force has vanishing contributions in both energy and helicity evolution equations (for the truncated system as well as for every non-linear triadic interaction).
Note that, because of the conservative dynamics of the lowest modes ${|\boldsymbol{k}|\leq k_F}$, the Euler forcing prevents the development of any inverse cascade.

If energy is concentrated at large scales in the initial spectrum, the transient dynamics of spectrally truncated 3D incompressible Euler equations behaves like dissipative Navier-Stokes equation and displays a K41 scaling \cite{bib:cichowlas2005}. However, we are interested here in the statistically stationary solution (statistical or absolute equilibrium).
The exact solutions for the statistical equilibrium energy and helicity spectra are~\cite{bib:kraichnan1973}
\begin{equation}
E(k)=\frac{8\pi}{\alpha}\frac{k^2}{1-\big(\frac{\beta}{\alpha}\big)^2k^2},\quad H(k)=\frac{8\pi\beta}{\alpha^2}\frac{k^4}{1-\big(\frac{\beta}{\alpha}\big)^2k^2},      \label{eq:exact_spectra}
\end{equation}
where $\alpha$ and $\beta$ depend on the total energy and helicity and are constrained by the realizability condition $|h(\boldsymbol{k})|\le 2\,|\boldsymbol{k}| \, e(\boldsymbol{k})$ such that $\alpha>0$ and $|\beta k_F|\le\alpha$.
Since, given a truncation wavenumber the solution depends only on the constant total energy and helicity, there is one independent non-dimensional parameter, \textit{e.g.} the relative helicity.

\begin{figure}[]
\unitlength 1mm
\begin{picture}(140,60)(0,0)
\put(5,0){\includegraphics[width=62\unitlength]{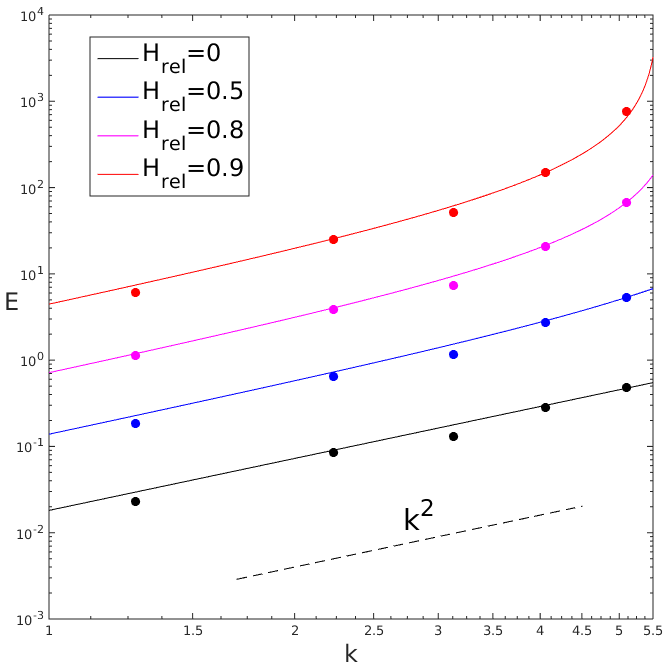}}
\put(80,0){\includegraphics[width=62\unitlength]{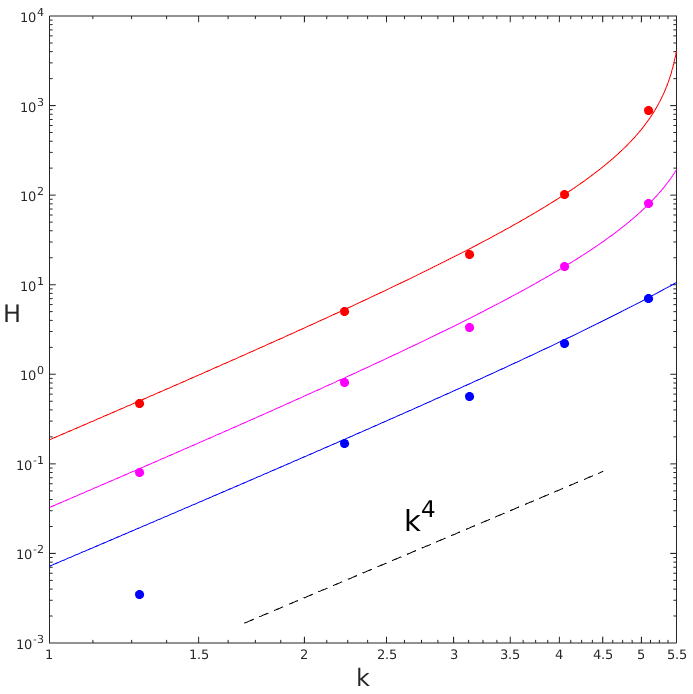}}
\put(37,55){(a)}
\put(110,55){(b)}
\end{picture}
        \caption{\label{fig:EHKraich} Exact energy and helicity spectra of the spherically truncated Euler system for different relative helicities (continuous line) and DNS time-averaged spectra (markers). The truncation wavenumber is $k_F=5.5$.} 
\end{figure}

Figure \ref{fig:EHKraich} shows the exact and numerical spectra for different relative helicities with $k_F=5.5$.
If $H_\text{rel}=0$ every wavevector holds the same amount of energy, and the energy spectrum is therefore proportional to $k^2$.
As the relative helicity increases, large wavenumber energy and helicity densities become larger and larger with respect to their low wavenumber counterparts.

Note that, with respect to previous works using the Euler forcing \cite{bib:naso2012, bib:pumir1994}, our implementation allows to control helicity injection and to vary $k_F$ arbitrarily, so that it is not restricted to non-helical turbulence and $k_F=1.5$.
In Euler-forced runs, Fourier coefficients for the forcing wavenumbers are initialized as a random solenoidal velocity field with a given energy spectrum.
In order to control the helicity injection  we have implemented both a non-helical and a helical modified initialization.
The mean helicity can be computed as $\sum_{\boldsymbol{k}} h(\boldsymbol{k})$, where the helicity density $h(\boldsymbol{k})= \boldsymbol{\hat{u}}(\boldsymbol{k})\cdot\boldsymbol{\hat{\omega}}^*(\boldsymbol{k})$ can be recast as  
$h(\boldsymbol{k})=2\, \boldsymbol{k} \cdot (\Re \boldsymbol{\hat{u}} \times \Im \boldsymbol{\hat{u}})$.
%
%
In helical Euler-forced simulations, the initial values of the forced modes are computed in order to obtain the maximal achievable helicity densities without changing the energy densities, \textit{i.e.}
\begin{equation}
\boldsymbol{\hat{u}}(\bk)=
e(\boldsymbol{k})^{1/2}\,\mathrm{e}^{\boldsymbol{\mathrm{i}}\,\gamma}\,\left(\boldsymbol{e}^{(1)}(\bk)+\mathrm{i}\boldsymbol{e}^{(2)}(\bk)\right)
\end{equation}
where $\gamma$ is a uniformly distributed random angle. 

In non-helical Euler-forced simulations, even if the initial velocity field described above already has nearly vanishing net helicity, we slightly modify the angles between the real and imaginary parts of all the forcing modes by the same quantity so that the net helicity is exactly zero,  \textit{i.e.} ${\sum_{\boldsymbol{k}} h(\boldsymbol{k})=0}$. 
Since the relative helicity in a helical forced simulation depends on the prescribed energy spectrum, we use different shapes for the initial energy spectrum in order to achieve different relative helicities.
The considered spectrum is $E(k)=k^pe^{-p/2(k/k_F)^2}$, with a maximum at $k=k_F$ and different possible values for $p$, \textit{e.g.} $p=4$ for a Batchelor spectrum and $p=2$ for a Saffman spectrum.
Our implementation of the truncated Euler equations has been validated against the spectra \eqref{eq:exact_spectra} predicted by Kraichnan \cite{bib:kraichnan1973}, as shown in figure  \ref{fig:EHKraich}.


\section{Anisotropy induced by forcing in non-rotating simulations}\label{sec:results1}
%
In this section, we study the anisotropy induced by the Euler, ABC and negative viscosity forcings on the statistics of non rotating (and expected to be isotropic) turbulence, namely energy, helicity and polarization angle-dependent spectra.
Except for the run forced through negative viscosity (for which only one velocity field is available), statistics of all the runs in this section and in section \ref{sec:results2} were obtained by time-averaging  over at least one eddy-turnover time after the statistically stationary state was reached. 
\begin{table}[] 
\centering
\begin{tabular}{l*{11}{c}r}
Run & Forcing & $k_F$ & $k_\text{max}\eta$ & $k_{\eta}$ & $Re^\lambda$ & $Re^L$ & ${H}_\text{rel}$ & Resolution\\[2pt]
\hline\\[-12pt]
A$_\text{nh}$ & non-hel. & 5.5 & 1.20 & 142 & 82.4 & 239 & -2.74E-3 & $512^3$ \\[2pt]
A$_\text{h}^1$ & helical & 5.5 & 1.22 & 140 & 81.3 & 219 & 0.451  & $512^3$ \\[2pt]
A$_\text{h}^2$ & helical & 5.5 & 1.19 & 143 & 81.7 & 210 & 0.617  & $512^3$ \\[2pt]
A$_\text{ABC}$ & ABC & 5 & 1.38 & 123 & 81.9 & 216 & 0.643  & $512^3$ \\[2pt]
B$_\text{h}$ & helical & 3.5 & 1.22 & 139 & 115 & 396 & 0.617  & $512^3$ \\[2pt]
B$_\text{ABC}$ & ABC & 3 & 1.45 & 117 & 116 & 397 & 0.622  & $512^3$ \\[2pt]
C$_\text{nh}^1$ & non-hel. & 1.5 & 2.71 & 62.6 & 110 & 432 & 9.61E-3  & $512^3$ \\[2pt]
C$_\text{nh}^2$ & non-hel. & 1.5 & 1.23 & 138 & 191 & 1208 & 7.38E-3  & $512^3$ \\[2pt]
C$_\text{h}^1$ & helical & 1.5 & 2.19 & 77.5 & 136 & 640 & 0.201  & $512^3$ \\[2pt]
C$_\text{h}^2$ & helical & 1.5 & 1.29 & 132 & 213 & 1469 & 0.227  & $512^3$ \\[2pt]
D$_\text{nv}$ & neg. visc. & 2.5 & 1.94 & 498 & 430 & 5587 & 8.22E-4 & $2048^3$  \\[2pt]
\end{tabular}
\caption{Parameters used in the non-rotating simulations: $k_\text{max}$ is the maximal resolved wavenumber (after dealiasing), $\eta$ is the Kolmogorov lengthscale and $k_{\eta}=1/\eta$.  $Re^\lambda$ and $Re^L$ are  Reynolds numbers respectively  based on the Taylor scale $\lambda$ and on the longitudinal integral lengthscale $L$. 
$H_\text{rel}$ refers to global relative helicity, \textit{i.e.} in Euler-forced runs it includes both the modes in the truncated system and the modes corresponding to wavenumbers outside the Euler sphere.
Letters A, B, C and D indicate different sets of non-rotating runs at decreasing $k_F$,  subscripts $_\text{nh}$, $_\text{h}$, $_\text{ABC}$ and $_\text{nv}$ stand for non-helical Euler, helical Euler, ABC and negative viscosity forcing, respectively. \label{tab:param}}
\end{table}
Table \ref{tab:param} reports the parameters of the non-rotating runs considered in this section.
We will show that anisotropy can be detected in most cases and that its characteristics depend on the forcing nature, on the value of the forcing wavenumber $k_F$, and on the relative helicity of the spherically truncated system in the case of helical Euler forcing.
Different Reynolds numbers are also considered.

For the runs in set A, $k_F=5.5$ (Euler forced runs) or $k_F=5$ (run A$_\text{ABC}$, ABC forcing). As a consequence, 738 modes are in the Euler sphere, while for run A$_\text{ABC}$ only 6 modes are involved in the forcing procedure (as in all ABC-forced runs). To allow a close comparison between all runs of the A series, we have ensured that the flow regimes are the same in terms of Reynolds numbers.
While in run A$_\text{nh}$ (non-helical Euler forced) the largest-wavenumber forcing modes contain the same energy as the lowest-wavenumber ones, in run A$_\text{h}^1$ (helical Euler forcing) the 48 largest-wavenumber modes (among 738 forcing modes) hold 15\% of the total energy.  
In comparison, in run A$_\text{h}^2$ (highly helical Euler forcing) the 48 largest-wavenumber modes hold 92\% of the kinetic energy in the Euler sphere and the relative helicity is nearly equal to that of run A$_\text{ABC}$.

We also perform simulations at a different forcing wavenumber $k_F$:
in set B, run B$_\text{h}$ is a helical Euler-forced run with {$k_F=3.5$} and large relative helicity, and run B$_\text{ABC}$ is an ABC-forced run with  {$k_F=3$}. 
Similarly to A$_\text{h}^2$, in run B$_\text{h}$ the 8 largest-wavenumber modes (among 178 forcing modes) hold 81\% of the Euler field energy, and the relative helicity is comparable to that of run B$_\text{ABC}$.

The non-helical and helical Euler forced runs in set C at $k_F=1.5$  allow to investigate the influence of the Reynolds number. $k_F=1.5$ is the lowest possible forcing wavenumber allowing non-linear interactions in the truncated system, which leads to 18 forcing modes.

Finally, run D$_\text{nv}$ is forced through the negative viscosity method and reaches the largest Reynolds number in the considered simulations, \textit{i.e.} $Re^\lambda=430$. Since $k_F=2.5$, 80 modes are forced. These data are provided by Kaneda's group \cite{bib:kanedajot}.
Only one instantaneous velocity field is available and ---in absence of time-averaging--- the resulting spectra are not as smooth as those from the other runs.

In the coming sections \ref{sec:isoenerg} to \ref{sec:isopol}, we investigate the statistics of forced turbulence, by measuring kinetic energy, helicity and polarization spectra, and examine the possible symmetry-breaking induced by the forcing by studying the dependence of these spectra on the polar angle.
\begin{figure}[]
\unitlength 1mm
\begin{picture}(140,50)(0,0)
\put(0,0){\includegraphics[width=70\unitlength]{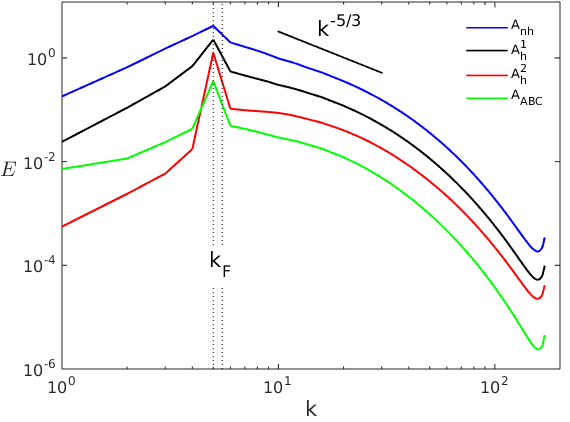}}
\put(75,0){\includegraphics[width=70\unitlength]{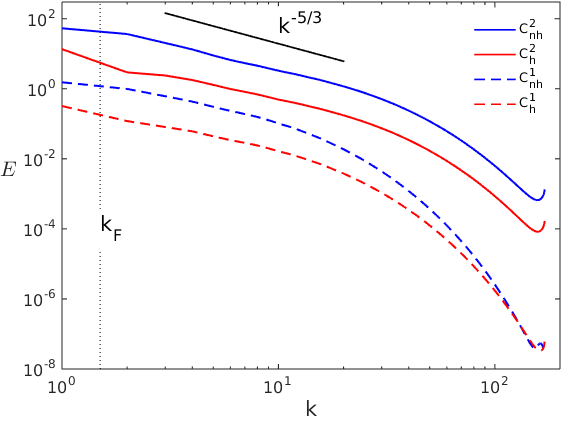}}
\put(35,10){(a)}
\put(110,10){(b)}
\end{picture}
        \caption{\label{fig:E1234}  Spherically integrated kinetic energy spectra: (a) for runs in set A (\mbox{$k_F=5$, $5.5$}); (b) for runs in set C ($k_F= 1.5$). Spectra are shifted with respect to each other, for better view. $k_F$ indicates the forcing wavenumber, as in the following plots.}
\end{figure}

\subsection{Energy spectra and energy directional anisotropy}
\label{sec:isoenerg}

Figure \ref{fig:E1234}  shows the spherically integrated kinetic energy spectra for runs in sets A and  C. The forcing wavenumber appears clearly as a marked peak in Fig.~\ref{fig:E1234}(a) ($k_F=5,5.5$).
When $k_F=1.5$, \textit{i.e.} in Fig.~\ref{fig:E1234}(b), since all the forcing wavevectors are included in the smallest shell, no energy peak is visible but a weak disturbance in the spectral slope around $k_F=1.5$ can be observed.
In this last figure, however, the Kolmogorov inertial scaling $k^{-5/3}$ appears in a wider spectral range than in Fig.~\ref{fig:E1234}(a) due to higher Reynolds numbers. In both figures, the presence of helicity in the forcing, and thereby of a helicity cascade, modifies the kinetic energy spectral scaling at wavenumbers slightly larger than $k_F$. In particular, in helical runs the energy spectra are flatter in a small range neighbouring $k_F$.

\begin{figure}[]
\unitlength 1mm
\begin{picture}(140,83)(0,0)
\put(0,0){\includegraphics[width=70\unitlength]{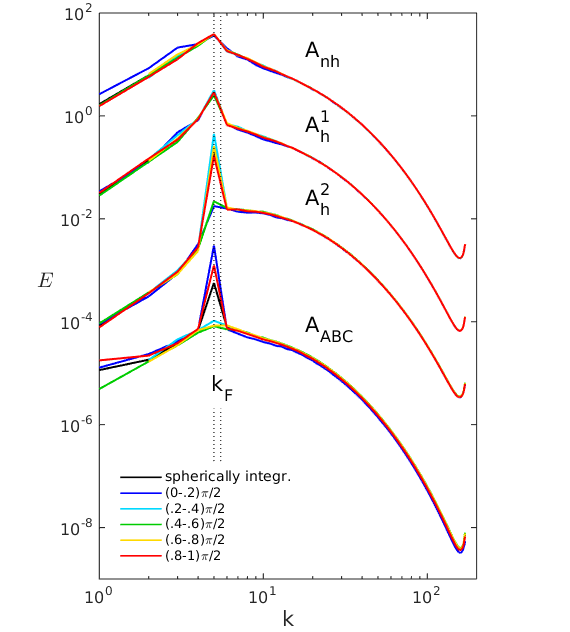}}
\put(75,0){\includegraphics[width=70\unitlength]{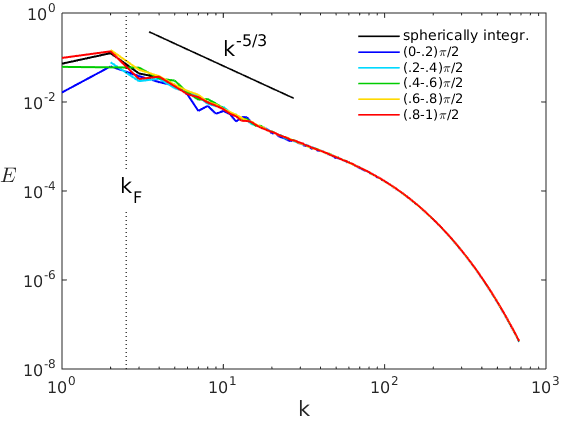}}
\put(40,10){(a)}
\put(110,10){(b)}
\end{picture}
         \caption{ \label{fig:spettrm1234} Directional energy spectra  $E_i(k)$ as functions of wavenumber $k$ for the five angular sectors for: (a) set A (\mbox{$k_F=5$, $5.5$}); (b) run D$_\text{nv}$.
The five sectors are indicated in legend, and the same colorcode applies throughout the paper.}
\end{figure}
%

Figure~\ref{fig:spettrm1234} shows the direction-dependent kinetic energy spectra $E_i(k)$ for runs of sets A and D. At first glance, over these logarithmic plots, the inertial and small scales are isotropic since all the curves at different orientations collapse on the spherically integrated spectrum $E(k)$, independently of the forcing method. Only in a vicinity of the forcing wavenumber, at large scales, does one observe a separation between the curves. This can be both attributed to less accurate sampling at low wavenumbers---although time-averages are used---and to the forcing.
Spectra of run A$_\text{ABC}$ (bottom set in Fig. \ref{fig:spettrm1234}(a)) seem to be more prone to this departure from isotropy over almost a decade of wavenumbers about the forcing one.

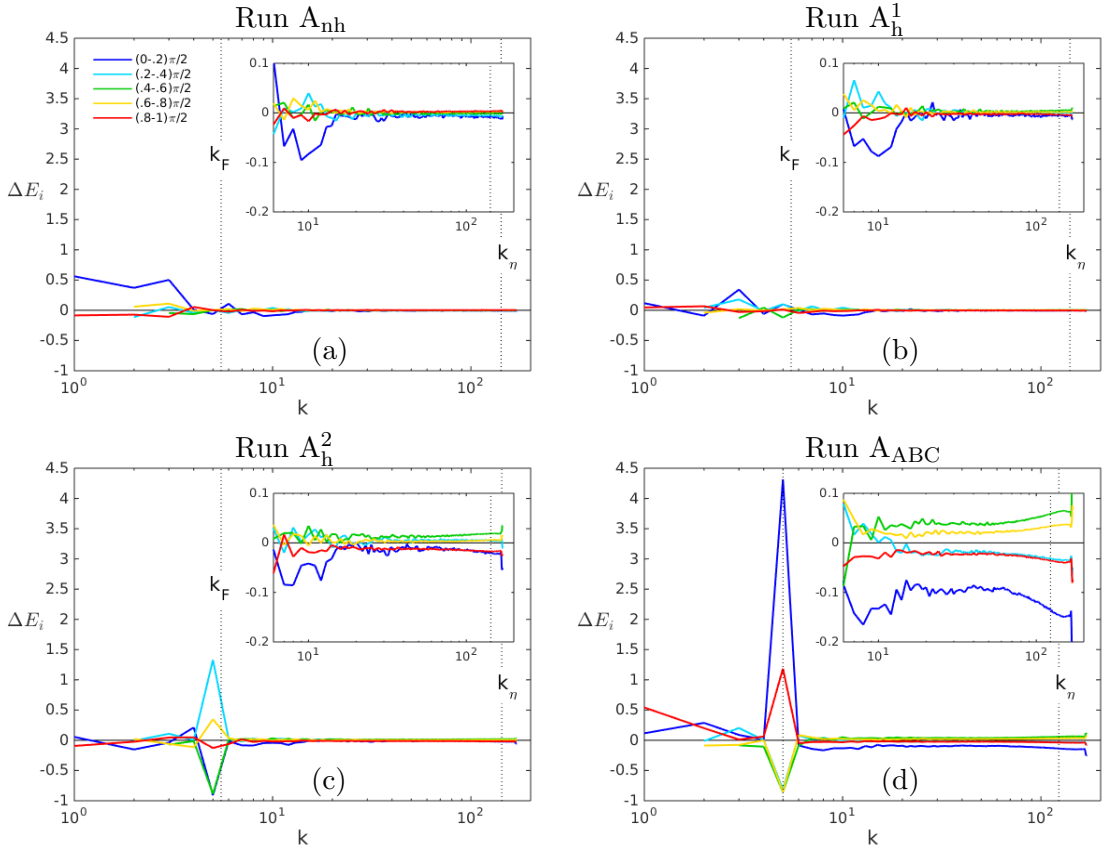
\begin{figure}[]
\unitlength 1mm
\begin{picture}(140,110)(0,0)
\put(0,57){\includegraphics[width=70\unitlength]{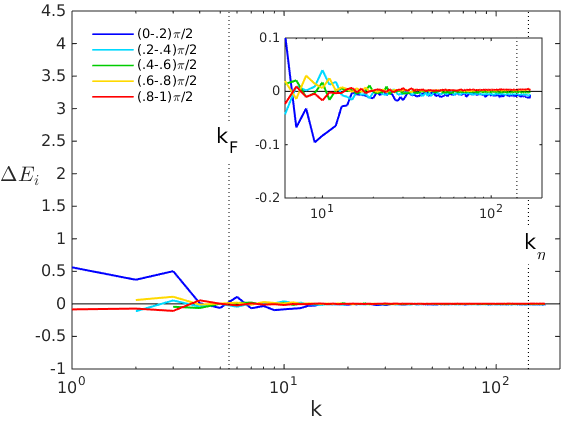}}
\put(75,57){\includegraphics[width=70\unitlength]{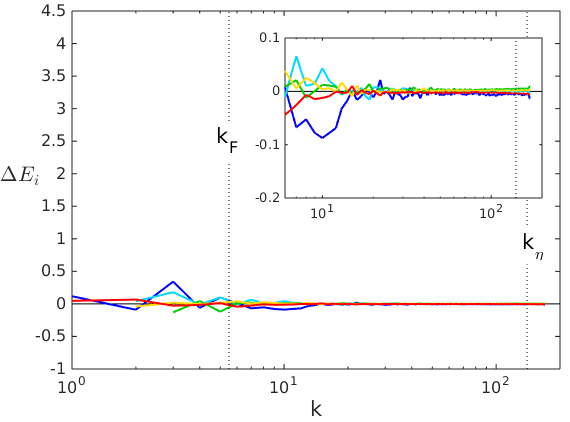}}
\put(0,0){\includegraphics[width=70\unitlength]{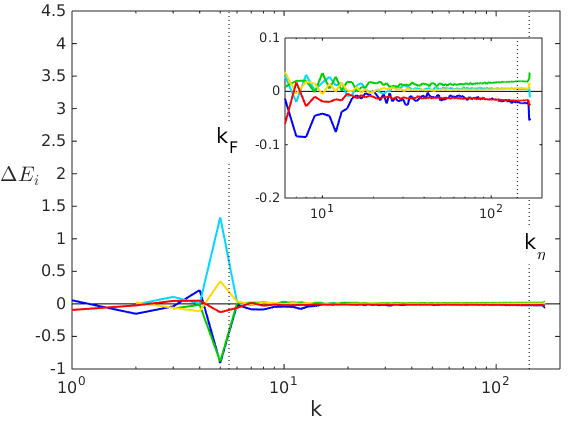}}
\put(75,0){\includegraphics[width=70\unitlength]{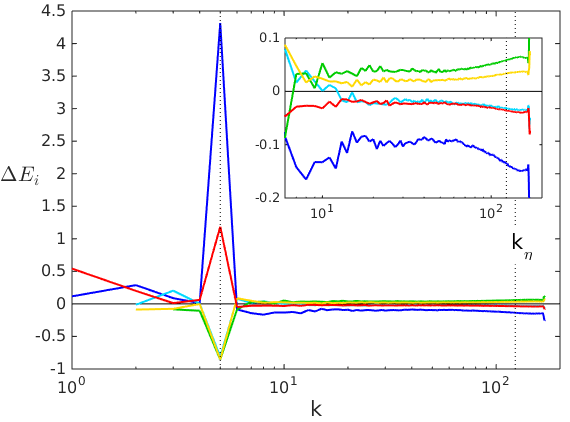}}
\put(40,65){(a)}
\put(115,65){(b)}
\put(40,8){(c)}
\put(115,8){(d)}
\put(30,109){Run A$_\text{nh}$}
\put(105,109){Run A$_\text{h}^1$}
\put(30,52){Run A$_\text{h}^2$}
\put(105,52){Run A$_\text{ABC}$}
\end{picture}
         \caption{\label{fig:direct_anis_1234} Directional anisotropy of the kinetic energy $\Delta E_i(k)$ for runs: (a)  A$_\text{nh}$; (b)  A$_\text{h}^1$; (c)  A$_\text{h}^2$; (d)  A$_\text{ABC}$. The insets focus on the large wavenumber inertial and dissipative ranges.}
\end{figure}

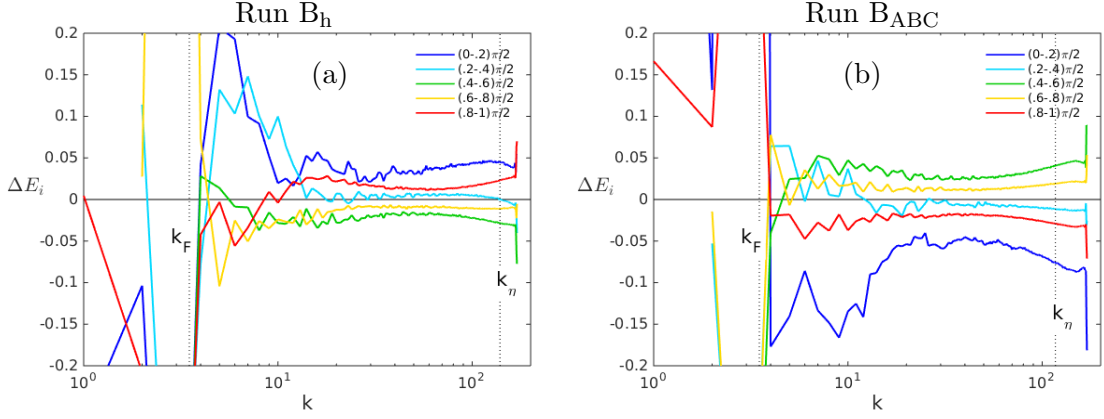
\begin{figure}[]
\unitlength 1mm
\begin{picture}(140,55)(0,0)
\put(0,0){\includegraphics[width=70\unitlength]{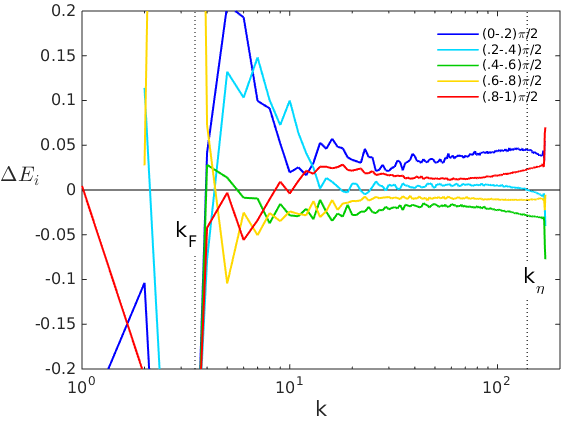}}
\put(75,0){\includegraphics[width=70\unitlength]{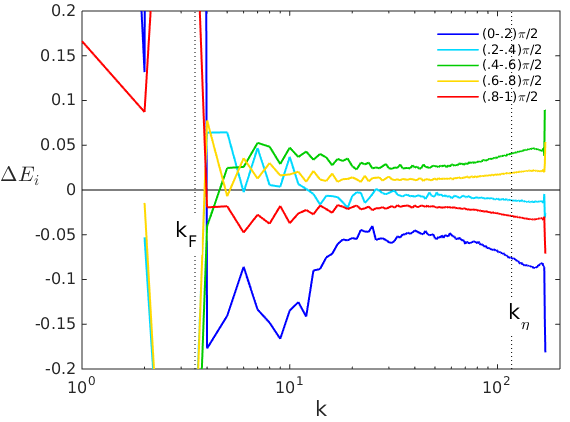}}
\put(40,44){(a)}
\put(110,44){(b)}
\put(30,52){Run B$_\text{h}$}
\put(105,52){Run B$_\text{ABC}$}
\end{picture}
         \caption{\label{fig:direct_anis_56} Directional anisotropy of the kinetic energy $\Delta E_i(k)$ for: (a) run B$_\text{h}$; (b) run B$_\text{ABC}$.} 
\end{figure}

\begin{figure}[]
\unitlength 1mm
\begin{picture}(140,55)(0,0)
\put(0,0){\includegraphics[width=70\unitlength]{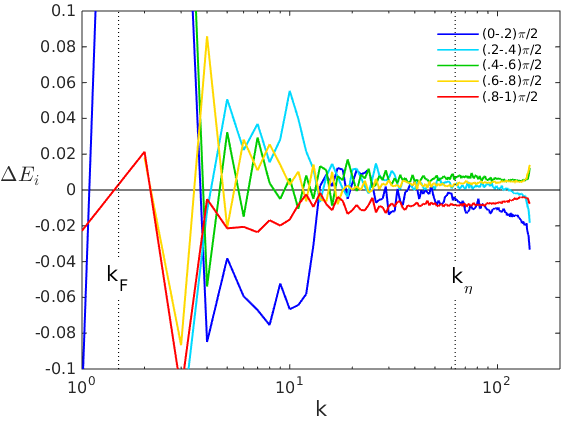}}
\put(75,0){\includegraphics[width=70\unitlength]{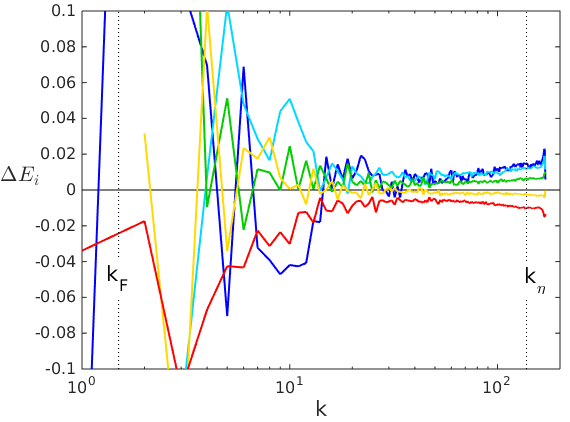}}
\put(40,44){(a)}
\put(110,44){(b)}
\put(30,52){Run C$_\text{nh}^1$}
\put(105,52){Run C$_\text{nh}^2$}
\end{picture}
         \caption{ \label{fig:direct_anis_78910} Directional anisotropy of kinetic energy $\Delta E_i(k)$ for: (a) run C$_\text{nh}^1$; (b) run C$_\text{nh}^2$.}
\end{figure}
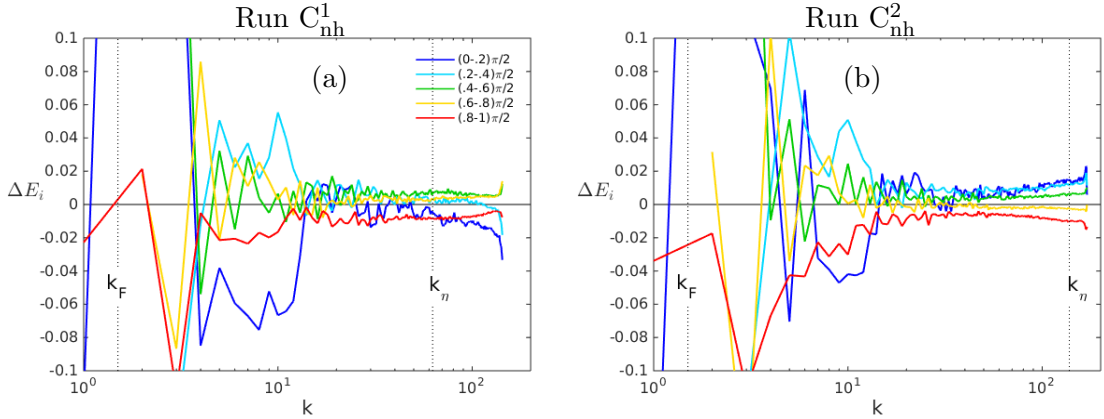

\begin{figure}
        \centering
\unitlength=1mm
                \includegraphics[width=70\unitlength]{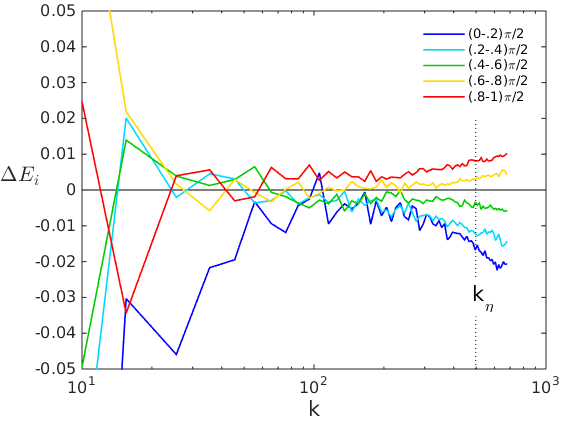}
%
         \caption{ \label{fig:direct_anis_Kaneda} Directional anisotropy of kinetic energy  $\Delta E_i(k)$ for run D$_\text{nv}$.}
\end{figure}

We however wish to focus more closely on the departure of the spectra from isotropy by investigating the relative difference between any directional spectrum and the spherically-integrated spectrum, computed as $\Delta E_i(k)=\left(E_i(k)-E(k)\right)/E(k)$  for $i=1,\cdots,5$.
This quantity is plotted in Figs. \ref{fig:direct_anis_1234} (set A), \ref{fig:direct_anis_56} (set B), \ref{fig:direct_anis_78910} (set C) and \ref{fig:direct_anis_Kaneda} (set D).
A quick observation of these figures shows that large-scale directional anisotropy develops in several runs. It is in particular confirmed that isotropy is generally better recovered in the small scales for the Euler-forced than for the ABC-forced runs, at the same Reynolds number and value of $k_F$.

In particular, one observes that:
\begin{enumerate}
\item for the same values of $k_F$ and $Re^\lambda$, the Euler-forced runs display an increasing anisotropy as their relative helicity increases (compare runs A$_\text{nh}$, A$_\text{h}^1$ and A$_\text{h}^2$ [Fig. \ref{fig:direct_anis_1234}(a)-(c)]);
\item for a similar level of helicity, the anisotropy of the Euler-forced runs is generally stronger at decreasing $k_F$ (compare runs A$_\text{nh}$ and C$_\text{nh}^1$ [Fig. \ref{fig:direct_anis_1234}(a) and \ref{fig:direct_anis_78910}(a)] for the non-helical case, or runs A$_\text{h}^2$ and B$_\text{h}$ for the helical case [Fig. \ref{fig:direct_anis_1234}(c) and \ref{fig:direct_anis_56}(a)]);
\item for similar values of $k_F$, ${H}_\text{rel}$ and $Re^\lambda$, the anisotropy is stronger in ABC-forced than in Euler-forced runs (compare runs A$_\text{h}^2$ and A$_\text{ABC}$ [Fig. \ref{fig:direct_anis_1234}(c) and \ref{fig:direct_anis_1234}(d)], or runs B$_\text{h}$ and B$_\text{ABC}$ [Fig. \ref{fig:direct_anis_56}(a) and \ref{fig:direct_anis_56}(b)]).
\end{enumerate}


All these results can be interpreted by considering the number of sufficiently excited modes in each run: the lower this number, the more anisotropy develops. This explains straightforwardly the aforementioned item 3. In fact the anisotropy level is the strongest in the ABC-forced runs since the ABC force excites directly only six modes: four in the horizontal sector and one in each vertical sector (see equation \eqref{eq:FABC2}). For runs A$_\text{ABC}$ (Fig. \ref{fig:direct_anis_1234}(d)) and B$_\text{ABC}$ (Fig. \ref{fig:direct_anis_56}(b)), the \mbox{$k_F$-centered} horizontal and vertical sectors hold more energy than the others. Nevertheless the opposite happens at small scales, which is consistent with the numerical  and theoretical results given by \cite{bib:Brasseur1987,1991tsf.....1...16B,bib:yeung1991,bib:yeung1995} for highly anisotropic forcing as recalled in the introduction.

Similarly, item 1 above can be explained by the fact that, when net helicity is large in the truncated Euler dynamics,  most of the energy remains concentrated in the largest wavenumbers so that only the corresponding modes are significantly excited by the forcing scheme. Therefore, if the number of the largest wavenumbers is sufficiently small, a small number of modes hold  most of the energy associated to the truncated Euler system and anisotropy develops.
However, the number of largest wavenumbers does not increase monotonically with $k_F$, and therefore a larger value of $k_F$ may yield larger small-scale anisotropy than a smaller value of $k_F$.
In fact, the anisotropy level of run B$_\text{h}$ (highly helical Euler forced, $k_F=3.5$, 8 largest wavevnumbers) is almost as large as the one of run B$_\text{ABC}$ (ABC forced, $k_F=3$), see Fig. \ref{fig:direct_anis_56}.
The opposite anisotropies for runs B$_\text{h}$ and run B$_\text{ABC}$ appearing in Fig. \ref{fig:direct_anis_56} depend on the different orientations of wavenumbers corresponding to the most excited modes.

We also observe in \textit{e.g.} Figs~\ref{fig:direct_anis_1234}(a) and \ref{fig:direct_anis_78910}(a)-(b), that non-helical Euler-forced simulations do not develop strong directional anisotropy. In fact, even the lowest possible forcing wavenumber allowing non-linear interactions in the truncated system, $k_F=1.5$, leads to 18 forcing modes, which have the same energy densities if the net helicity of the truncated Euler system is zero.
Furthermore, the anisotropy level of a $k_F=1.5$ helical Euler forced run cannot be as strong as that in run B$_\text{h}$ ($k_F=3.5$, 8 largest wavenumbers) or in runs A$_\text{ABC}$ and B$_\text{ABC}$ (ABC-forced, 6 forcing modes), because in the sphere of radius $k_F=1.5$ there are 12 largest wavenumbers (the ones with two unitary components and one null component).\\

We finally investigate the influence of the Reynolds number. Figure \ref{fig:direct_anis_78910} shows the energy directional anisotropy for the non-helical runs in set C (Euler forced, $k_F=1.5$, at moderate and high Reynolds numbers). The results for the helical runs C$_\text{h}^1$ and C$_\text{h}^2$, not shown here, are qualitatively similar.
By comparing the moderate Reynolds number case in run C$_\text{nh}^1$ (Fig. \ref{fig:direct_anis_78910}(a)) with the higher Reynolds number case in run C$_\text{nh}^2$ (Fig. \ref{fig:direct_anis_78910}(b)), no obvious trend towards isotropy is observed at increasing wavenumber and Reynolds number.
Instead, for the largest Reynolds number, anisotropy clearly increases with the wavenumber, in agreement with \cite{bib:Brasseur1987,1991tsf.....1...16B,bib:yeung1991,bib:yeung1995}.
The same behavior is observed in run D$_\text{nv}$ at an even higher Reynolds number, as shown in Fig. \ref{fig:direct_anis_Kaneda} (note that the spectra plotted in this figure have been obtained by using larger bins than in the other cases, since no time-averaging is possible over this single velocity field snapshot).

\begin{figure}[]
\unitlength 1mm
\begin{picture}(140,80)(0,0)
\put(0,0){\includegraphics[width=70\unitlength]{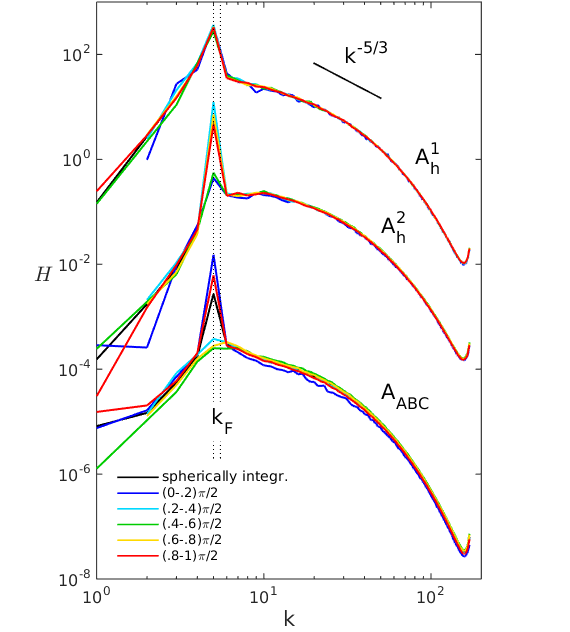}}
\put(70,0){\includegraphics[width=70\unitlength]{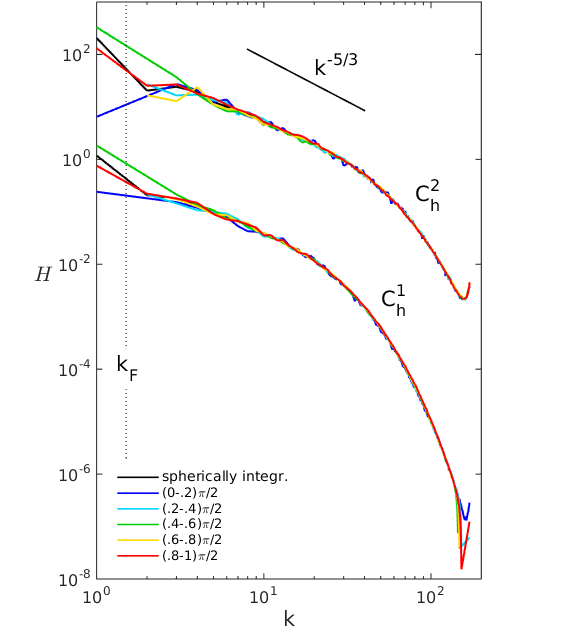}}
\put(40,9){(a)}
\put(110,9){(b)}
\end{picture}
         \caption{\label{fig:direct_anisH_1234} Directional helicity spectra $H_i(k)$ for: (a) runs A$_\text{h}^1$, A$_\text{h}^2$ and A$_\text{ABC}$; (b) runs C$_\text{h}^2$ and C$_\text{h}^1$.}
\end{figure}

\subsection{Helicity spectra and helicity directional anisotropy}
%

%
\begin{figure}[]
\unitlength 1mm
\begin{picture}(140,50)(0,0)
\put(0,0){\includegraphics[width=70\unitlength]{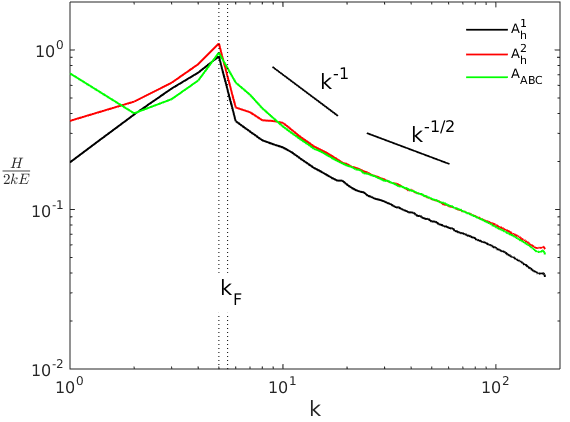}}
\put(75,0){\includegraphics[width=70\unitlength]{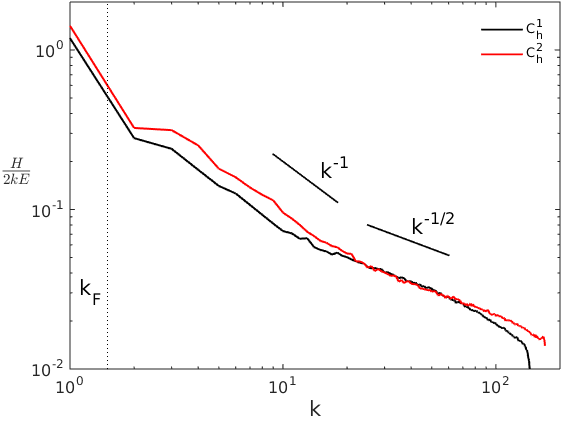}}
\put(35,8){(a)}
\put(110,8){(b)}
\end{picture}
         \caption{\label{fig:hrel234}Relative helicity spectra $H(k)/(2kE(k))$ for helical runs in: (a) set A; (b) set C.}
\end{figure}
Figure~\ref{fig:direct_anisH_1234}  shows helicity directional spectra for helical runs in sets A and C. Like in energy directional spectra, no small scale anisotropy can be detected from these helicity spectra, and large scales seem to be more anisotropic in the case of ABC forcing. By comparing run A$_\text{h}^2$ with run A$_\text{ABC}$ in Fig.~\ref{fig:direct_anisH_1234}(a), one can observe that the ABC-forced run displays a wider inertial range, even though these two runs have similar Reynolds numbers.
Figure \ref{fig:hrel234}  shows relative helicity spectra $H(k)/(2kE(k))$ for helical runs in sets A and C.
A slope close to $k^{-1}$ at low wavenumbers indicates that energy and helicity spectra scale with the same power of $k$ at large scales. The small-scale $-1/2$ slope has already been reported in previous studies of both isotropic \cite{bib:mininni_alexakis_pouquet_2008} and rotating \cite{bib:mininni2012} helical turbulence. The maximal value of relative helicity is approximately $1$ and is obtained in the shells containing wavenumbers with modulus $k_F$ for all five simulations.

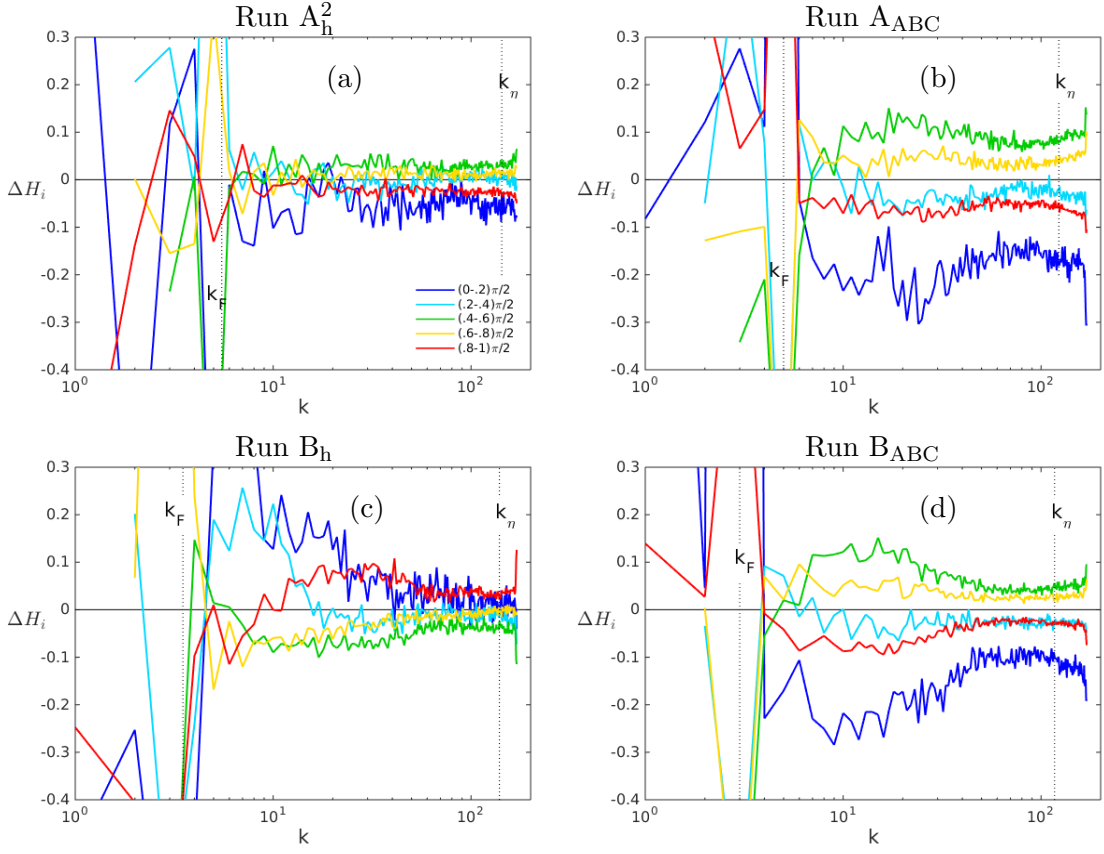
\begin{figure}[]
\unitlength 1mm
\begin{picture}(140,110)(0,0)
\put(0,57){\includegraphics[width=70\unitlength]{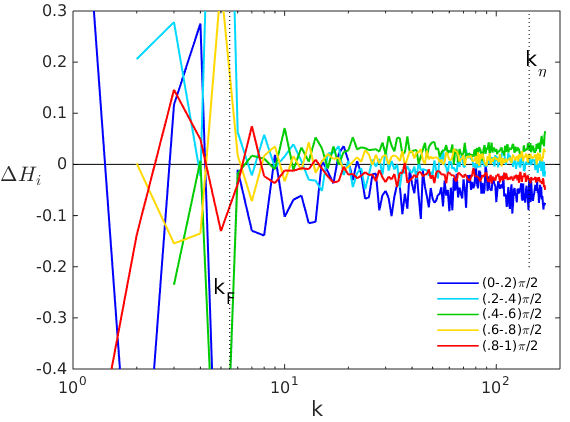}}
\put(75,57){\includegraphics[width=70\unitlength]{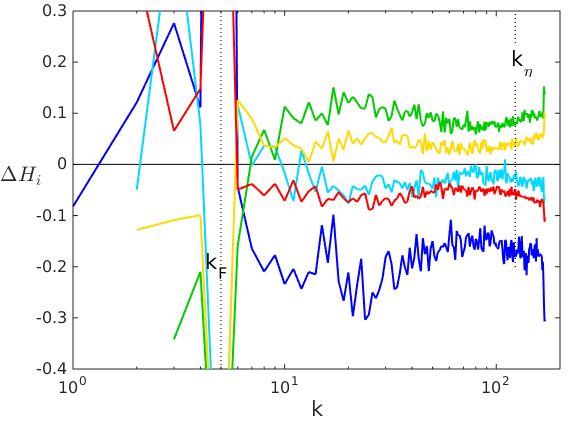}}
\put(0,0){\includegraphics[width=70\unitlength]{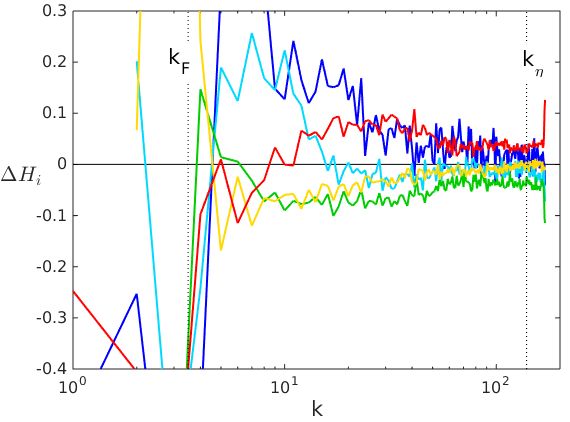}}
\put(75,0){\includegraphics[width=70\unitlength]{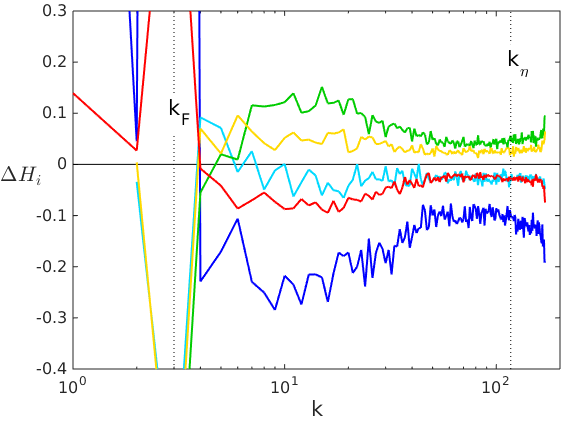}}
\put(42,101){(a)}
\put(120,101){(b)}
\put(45,44){(c)}
\put(120,44){(d)}
\put(30,109){Run A$_\text{h}^2$}
\put(105,109){Run A$_\text{ABC}$}
\put(30,52){Run B$_\text{h}$}
\put(105,52){Run B$_\text{ABC}$}
\end{picture}
         \caption{\label{fig:direct_anisH_3456} Helicity directional anisotropy $\Delta H_i(k,\theta)$ for: (a) run A$_\text{h}^2$; (b) run A$_\text{ABC}$; (c) run B$_\text{h}$; (d) run B$_\text{ABC}$.}
\end{figure}

As for the kinetic energy, we define the normalised departure of the directional helicity spectrum from the spherically-integrated one as $\Delta H_i(k)=(H_i(k)-H(k))/H(k)$.
Figure \ref{fig:direct_anisH_3456} shows helicity directional anisotropy for some helical runs in sets A and B. 
The distribution of directional anisotropy is similar between energy and helicity (compare Fig. \ref{fig:direct_anisH_3456}(a,b) with Fig. \ref{fig:direct_anis_1234}(c,d), and Fig. \ref{fig:direct_anisH_3456}(c,d) with Fig. \ref{fig:direct_anis_56}(a,b)). Conclusions similar to those presented in section~\ref{sec:isoenerg} can therefore be drawn for helicity, that is, the ABC-forced runs display a higher level of directional anisotropy with respect to Euler-forced runs.

In summary, by looking at the results obtained for the most anisotropic forcings, that is Figs. \ref{fig:direct_anis_1234}(d), \ref{fig:direct_anis_56}(b), \ref{fig:direct_anisH_3456}(b), and \ref{fig:direct_anisH_3456}(d), that represent energy and helicity directional anisotropy for the ABC-forced runs with $k_F=5$ and $k_F=3$, and Fig. \ref{fig:direct_anis_56}(a) (directional anisotropy of the highly helical Euler-forced run with $k_F=3.5$), it is clear that the anisotropy for each angular sector is constant down to the smallest resolved scales or that it even increases with the wavenumber. This is consistent with the results of Yeung \& Brasseur \cite{bib:yeung1991,bib:yeung1995} for highly anisotropic forcings.


\begin{figure}[]
\unitlength 1mm
\begin{picture}(140,110)(0,0)
\put(0,57){\includegraphics[width=70\unitlength]{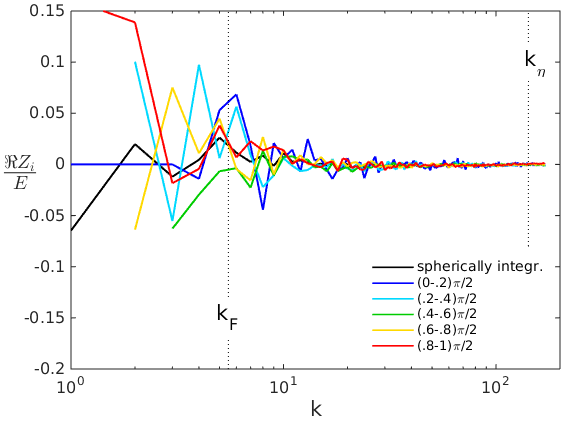}}
\put(75,57){\includegraphics[width=70\unitlength]{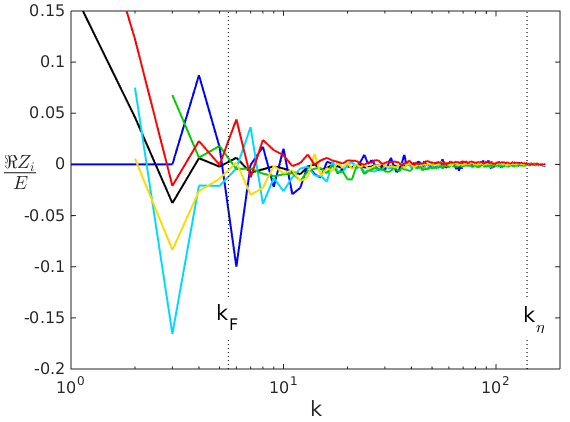}}
\put(0,0){\includegraphics[width=70\unitlength]{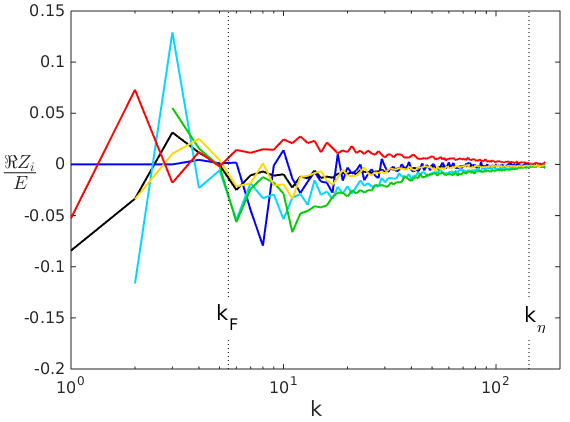}}
\put(75,0){\includegraphics[width=70\unitlength]{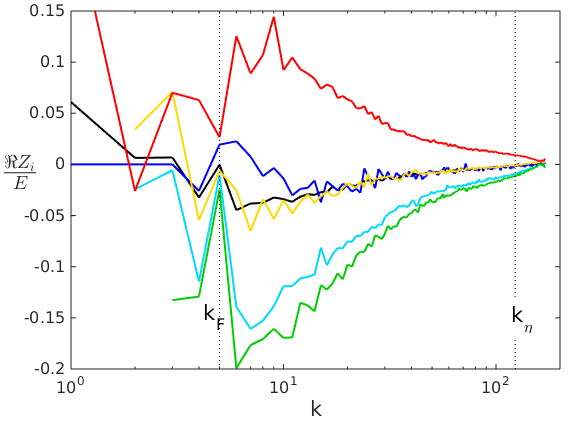}}
\put(40,67){(a)}
\put(115,67){(b)}
\put(40,10){(c)}
\put(115,10){(d)}
\put(30,109){Run A$_\text{nh}$}
\put(105,109){Run A$_\text{h}^1$}
\put(30,52){Run A$_\text{h}^2$}
\put(105,52){Run A$_\text{ABC}$}
\end{picture}
         \caption{\label{fig:diffE21_1234} Normalised directional polarization spectra $\Re Z_i(k)/E(k)$ for: (a) run A$_\text{nh}$; (b) run A$_\text{h}^1$; (c) run A$_\text{h}^2$; (d) run A$_\text{ABC}$.}
\end{figure}
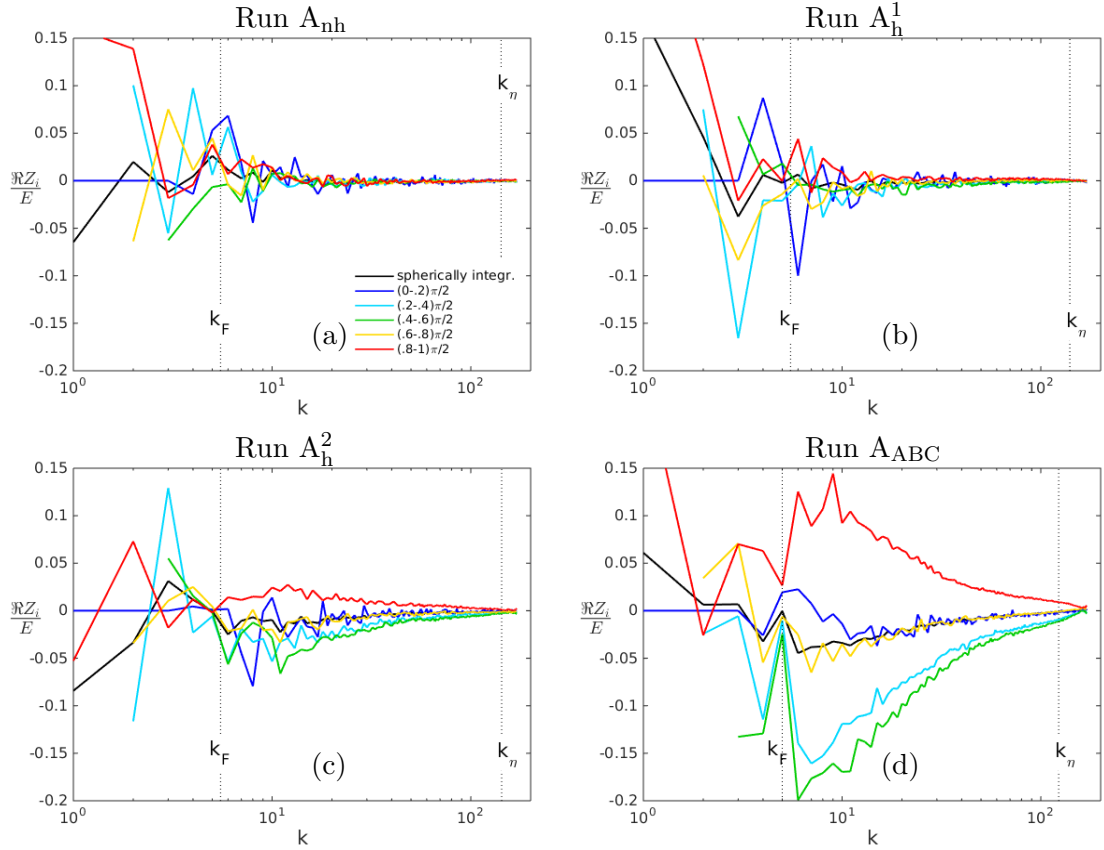

\subsection{Polarization anisotropy}
\label{sec:isopol}

We focus now on the directional dependence of $z(\bk)$ through the normalised spectrum and the normalised directional spectra of its real part,  
$\Re{Z}(k)/E(k)=(E^\text{pol}(k)-E^\text{tor}(k))/E(k)$ and $\Re{Z_i}(k)/E(k)=(E^\text{pol}_i(k)-E^\text{tor}_i(k))/E(k)$.  We recall that $\Re{Z}(k)=0$ in strictly isotropic turbulence.

The $\Re{Z}(k)/E(k)$ quantity is plotted in Fig. \ref{fig:diffE21_1234} for runs of set A (the other runs, not shown, display similar trends).
In the non-helical run (Fig. \ref{fig:diffE21_1234}(a)), $\Re{Z}(k)$ displays the features expected in strictly isotropic turbulence (that is, its spherically integrated spectrum and its directional spectra vanish up to statistical uncertainty), both in the inertial and in the dissipative ranges.
Runs A$_\text{h}^1$ (Fig. \ref{fig:diffE21_1234}(b)) and A$_\text{h}^2$ (Fig. \ref{fig:diffE21_1234}(c)) clearly show that, in Euler forced runs, the presence of helicity induces a slight polarization anisotropy over most of the inertial and dissipative ranges, and that isotropy is recovered only at the smallest scales.
Larger values of the polarization anisotropy are found in the inertial range in ABC-forced (Fig. \ref{fig:diffE21_1234}(d)) and highly helical Euler-forced runs (Fig. \ref{fig:diffE21_1234}(c)) due to the relatively low number of excited modes, but it is definitely larger in the ABC-forced case.


\section{Anisotropy induced by rotation}
\label{sec:results2}
In the previous section we studied the anisotropy artificially induced by forcing, which is a necessary pre-requisite before assessing the 
global anisotropic structure of forced rotating turbulence. 
In the present section, we consider rotating homogeneous turbulence which we simulate numerically as in the previous section  but setting $\bOmega\ne {\bf 0}$ in Eq. (\ref{eq:NS}).
In the latter context, anisotropy
has two contributions: one, artificial, due to the forcing (as illustrated in section \ref{sec:results1}), and another one inherent to the phenomenology of rotating flows \textit{per se}.
However---unlike in the non-rotating case---in the presence of background rotation, the Euler and ABC forcing schemes \textit{a priori} give rise to substantially different physical systems. Indeed, as already explained in Section \ref{sec:forc}, while in Euler-forced runs the modes in the spherically truncated system evolve independently of the other modes, the low wavevectors modes in ABC-forced runs are coupled with all the other modes. As a consequence, if rotation is large enough, energy is allowed to cascade backward. This inverse cascade, previously observed in \cite{bib:mininnicascade}, manifests as an increase of the energy in the smallest wavenumbers, and as a consequence the flow is not statistically stationary.

In this section we first investigate the effect of rotation and of helicity, and the differences of anisotropy between ABC-forced runs, with a dual (direct and inverse) cascade, and helical Euler-forced runs, with only forward cascade. Then we study through high resolution Euler-forced runs the anisotropy that naturally arises because of background rotation in the absence of inverse cascade. Both the rotation rate vector and the fixed direction $\boldsymbol{n}$ defining the Craya frame are in the $x_3$ direction.
Table \ref{tab:param2} reports the parameters of $512^3$ (set R) and $1024^3$ (set S) rotating runs forced through Euler and ABC schemes.
In all Euler-forced runs presented in this table, the spherically truncated Euler equation includes the Coriolis force. We also performed runs without rotation in the Euler system and observed no significant change in the small scale anisotropy.
\begin{table}[] 
\centering
\begin{tabular}{l*{11}{c}r}
Run & Forcing & $k_F$ & $k_\text{max}\eta$ & $k_{\eta}$ & $k_{\Omega}$ & $Re^\lambda$ & $Re^L$ & $Ro^\omega$ & $Ro^L$ & ${H}_\text{rel}$ & Resolution \\[2pt]
\hline\\[-12pt]
R$_\text{nh}^1$ & non-hel. & 5.5 & 1.21 & 140 & 43.4 & 111 & 373 & 1.26 & 0.206 & -6.60E-3  & $512^3$ \\[2pt]
R$_\text{nh}^2$ & non-hel. & 5.5 & 1.14 & 149 & 82.5 & 149 & 435 & 0.857 & 0.161 & -2.70E-3 & $512^3$ \\[2pt]
R$_\text{h}$ & helical & 5.5 & 1.34 & 127 & 48.0 & 116 & 307 & 1.10 & 0.228 & 0.522  & $512^3$\\[2pt]
R$_\text{ABC}$ & ABC & 5 & 2.42 & 70.2 & 25.7 & 111 & 351 & 1.13 & 0.195 & 0.591 & $512^3$ \\[2pt]
S$_\text{nh}^1$ & non-hel. & 5.5 & 1.16 & 295 & 9.01 & 151 & 808 & 5.91 & 0.605 & -2.85E-3  & $1024^3$\\[2pt]
S$_\text{nh}^2$ & non-hel. & 5.5 & 1.17 & 290 & 44.1 & 187 & 959 & 2.03 & 0.216 & -5.27E-3 & $1024^3$ \\[2pt]
S$_\text{h}$ & helical & 5.5 & 1.29 & 264 & 47.5 & 193 & 797 & 1.81 & 0.240 & 0.386  & $1024^3$\\[2pt]
\end{tabular}
\caption{\label{tab:param2} Parameters used in the rotating turbulence simulations.
$k_\Omega=\left(\left(2\Omega\right)^3/\epsilon\right)^{1/2}$ is the Zeman wavenumber.
Letters R and S refer to runs at resolutions $512^3$ and $1024^3$, respectively. The other definitions are the same as in Tab. \ref{tab:param}.}
\end{table}

\subsection{Effects of forcing anisotropy and  inverse cascade}
\begin{figure}
\centering
\unitlength 1mm
\includegraphics[width=70\unitlength]{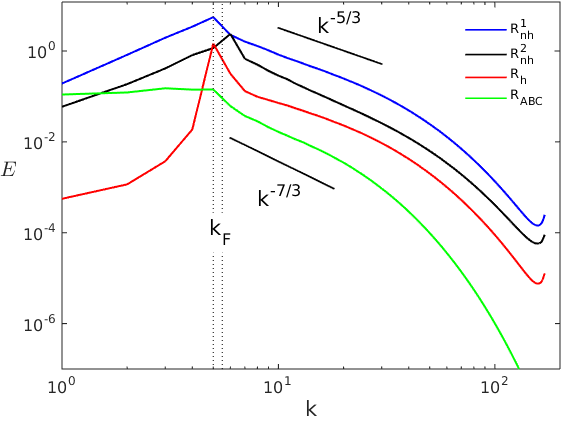}
         \caption{\label{fig:Ehv}Spherically averaged energy spectra $E(k)$, for runs in set R.}
\end{figure}

\begin{figure}
        \centering
\unitlength 1mm
     \includegraphics[width=70\unitlength]{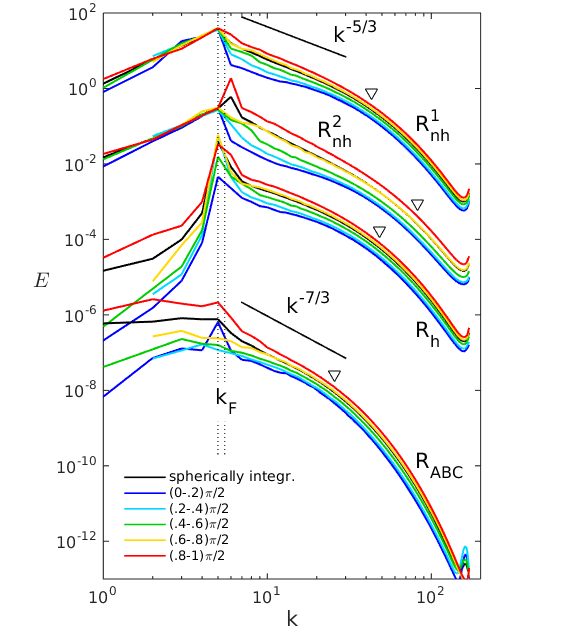}
                \caption{\label{fig:spettrmrot} Directional energy spectra $E_i(k)$ for runs in set R. Markers indicate the Zeman scale \mbox{$k_\Omega=\sqrt{\left(2\Omega\right)^3/\epsilon}$}.}
\end{figure}

\begin{figure}[]
\unitlength 1mm
\begin{picture}(140,110)(0,0)
\put(0,57){ \includegraphics[width=70\unitlength]{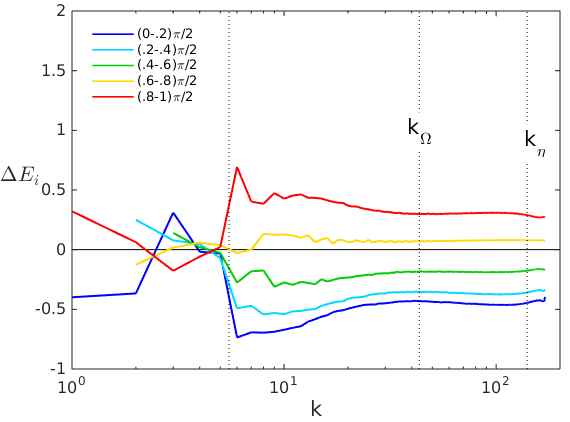}}
\put(75,57){\includegraphics[width=70\unitlength]{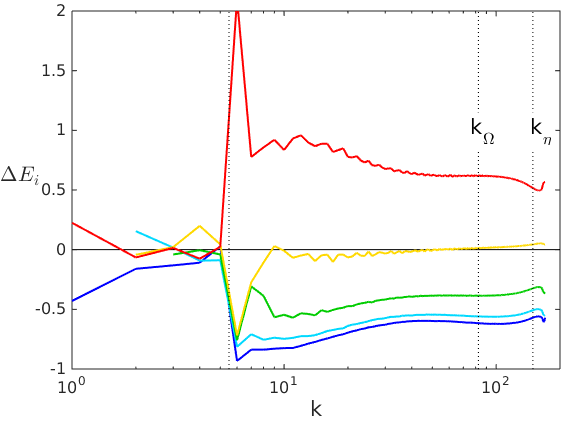}}
\put(0,0){ \includegraphics[width=70\unitlength]{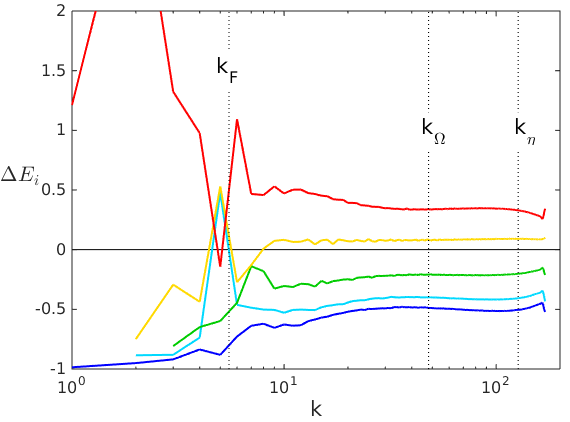}}
\put(75,0){\includegraphics[width=70\unitlength]{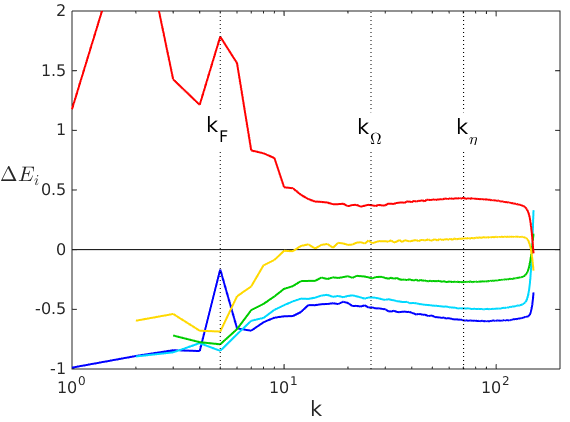}}
\put(35,101){(a)}
\put(110,101){(b)}
\put(35,44){(c)}
\put(110,44){(d)}
\put(30,109){Run R$_\text{nh}^1$}
\put(105,109){Run R$_\text{nh}^2$}
\put(30,52){Run R$_\text{h}$}
\put(105,52){Run R$_\text{ABC}$}
\end{picture}
          \caption{\label{fig:direct_anis_1316} Relative directional anisotropy of kinetic energy $\Delta E_i(k)$ for: (a) run R$_\text{nh}^1$; (b) run R$_\text{nh}^2$; (c) run R$_\text{h}$; (d) run R$_\text{ABC}$.}
\end{figure}

As shown in Table~\ref{tab:param2}, runs R$_\text{nh}^1$, R$_\text{h}$ and R$_\text{ABC}$ have comparable Reynolds and Rossby numbers, runs R$_\text{h}$ and R$_\text{ABC}$ also have comparable relative helicity, and run R$_\text{nh}^2$ has a Rossby number significantly lower than the other three runs. Thus, by comparing run R$_\text{h}$ with run R$_\text{ABC}$ one can estimate the combined effect of the forcing nature and of the presence of an inverse cascade, when turbulence is subject to a background rotation. Furthermore, the comparison of runs R$_\text{nh}^1$ and R$_\text{h}$ permits to study the effect of helicity, and comparing run R$_\text{nh}^1$ with run R$_\text{nh}^2$ allows to study the effect of a decrease in Rossby number.
Figures \ref{fig:Ehv}, \ref{fig:spettrmrot} and \ref{fig:direct_anis_1316} show the spherically averaged energy spectra, directional energy spectra and energy  directional anisotropy for runs in set R.

Figure~\ref{fig:Ehv} shows that, in runs R$_\text{nh}^1$ (moderate Rossby number non-helical Euler forced) and R$_\text{h}$ (moderate Rossby number helical Euler forced) the slope of the energy spectrum is close to $-5/3$.
Since the Reynolds number is not very large, this is a consequence of weak rotation, as argued and observed in DNS by \cite{bib:binbaqui2015}.
However run R$_\text{ABC}$ (moderate Rossby number ABC-forced), which has Reynolds and Rossby numbers values comparable to those of the Euler forced runs, shows a steeper spectrum ($-7/3$ slope), rather close to the one of run R$_\text{nh}^2$ (low Rossby number non-helical Euler-forced).
Note that a slope equal to $-2.2$ was already observed in \cite{bib:mininni2009,bib:mininni2012} for rotating DNS forced through the ABC forcing.

Figure \ref{fig:spettrmrot} shows the direction-dependent kinetic energy spectra for the same runs.
From this energetic point of view, wavevectors closer to the horizontal plane $\bk \cdot \bOmega=0$ (red curves) hold more energy than wavevectors closer to $\bOmega$ (blue curves), thereby indicating a trend towards two-dimensionalisation as expected in the presence of rotation.
Directional anisotropy is larger at large scales than at small scales, as shown by the departure between the less energetic vertical orientation ($\theta\simeq 0$) and the more energetic horizontal orientation ($\theta\simeq \pi/2$) of wavevector. 
However, the small scales are still significantly anisotropic. In fact, for runs in set R, the Zeman wavenumber $k_\Omega$ is relatively large: observing anisotropy at all scales is therefore consistent with the classical dimensional argument according to which isotropy should be recovered only at scales significantly smaller than the Zeman scale \cite{bib:mininni2012,bib:delache2014,bib:zeman}.

Second, considering only the relative anisotropy in the energy spectrum, we compute the scale-normalised departure between each directional spectrum and the corresponding average spectrum, $\Delta E_i(k)=(E_i(k)-E(k))/E(k)$.
Figure~\ref{fig:direct_anis_1316} shows this quantity for the four runs of set R. It  confirms that the relative anisotropy persists through the inertial scales down to the smallest ones, and that the difference between $E_1(k)$ and $E_5(k)$ is up to $100$\% for the strongly rotating non helical Euler-forced flow (run $R^2_\text{nh}$, Fig. \ref{fig:direct_anis_1316}(b)), and $50\%$ for the other runs. 
The energy directional anisotropy inherently induced by the ABC forcing and evidenced in Section \ref{sec:results1} is partly concealed in the anisotropy due to large rotation, as observed when comparing Figs. \ref{fig:direct_anis_1316}(c) (helical Euler forced) and \ref{fig:direct_anis_1316}(d) (ABC forced).
The effect of helicity can be deduced by comparing Figs. \ref{fig:direct_anis_1316}(a) and \ref{fig:direct_anis_1316}(c), which shows that the presence of helicity has no significant effect on small scale  anisotropy.
Note finally that the presence of the inverse cascade has no clear effect on small-scale energy directional anisotropy (compare Fig. \ref{fig:direct_anis_1316}(c) and (d)).


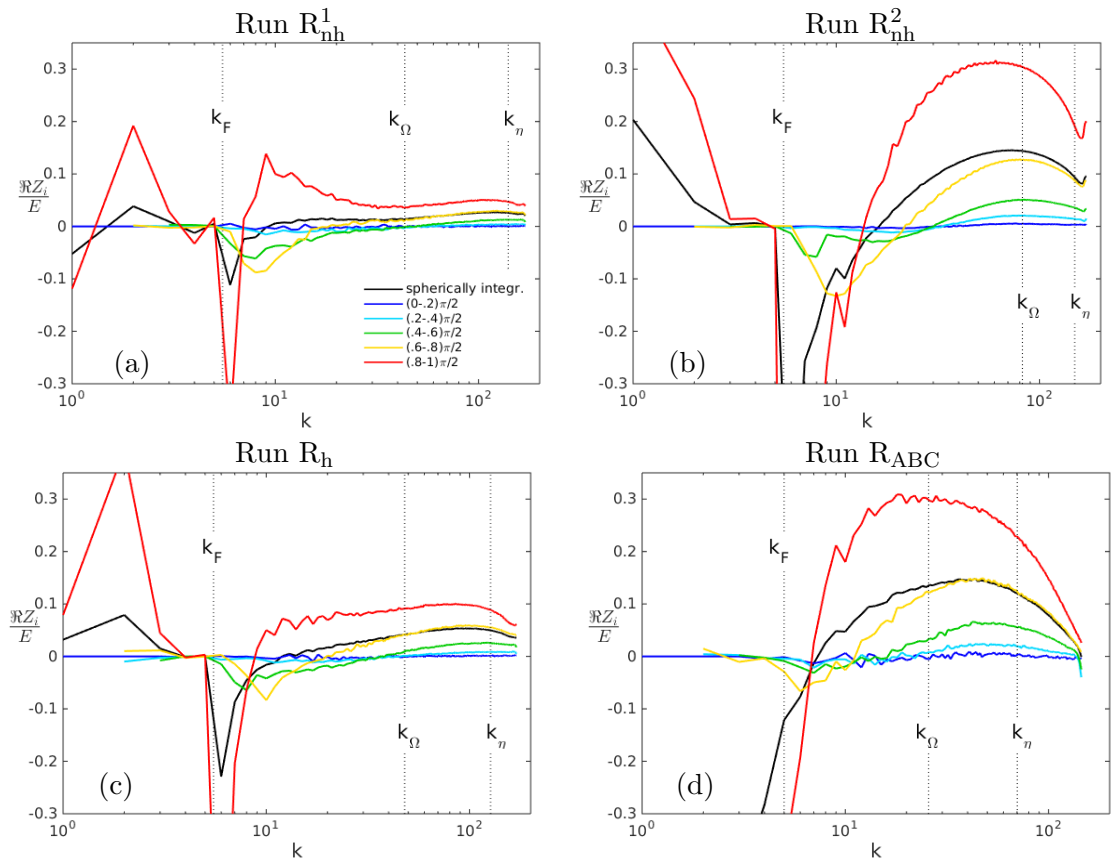
\begin{figure}[]
\unitlength 1mm
\begin{picture}(140,110)(0,0)
\put(0,57){ \includegraphics[width=70\unitlength]{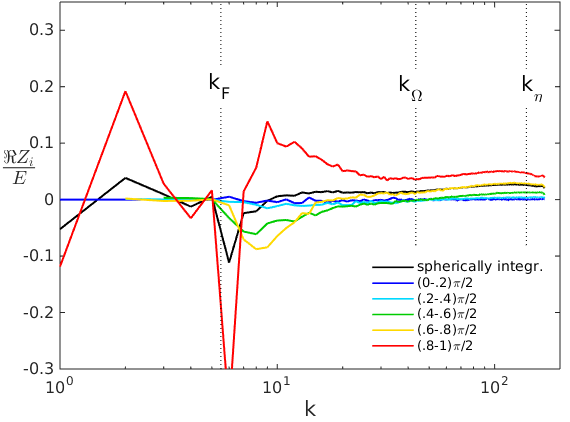}}
\put(75,57){\includegraphics[width=70\unitlength]{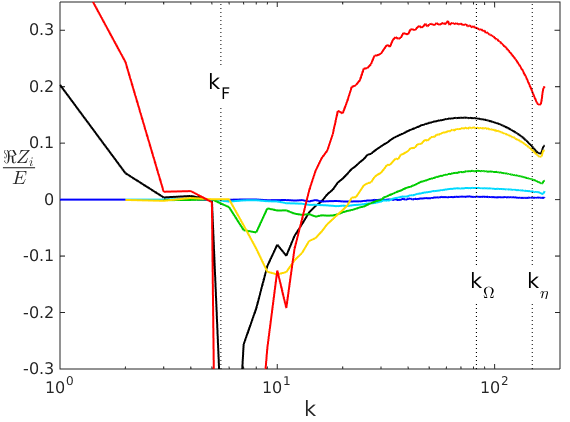}}
\put(0,0){\includegraphics[width=70\unitlength]{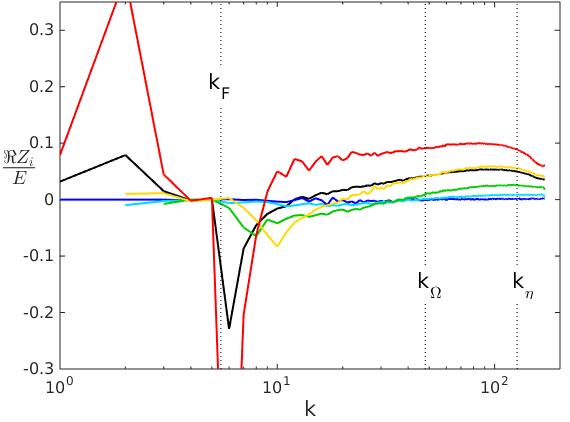}}
\put(75,0){ \includegraphics[width=70\unitlength]{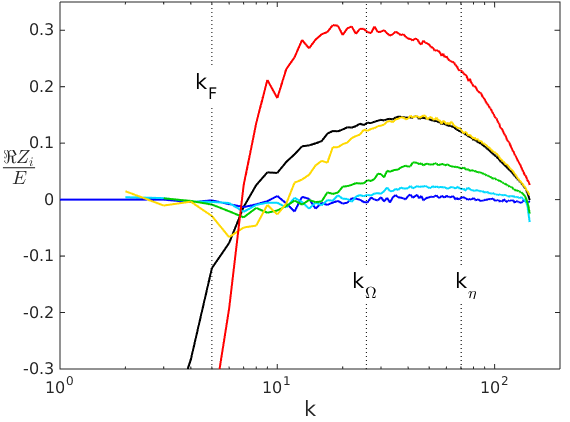}}
\put(30,110){Run R$_\text{nh}^1$}
\put(105,110){Run R$_\text{nh}^2$}
\put(30,53){Run R$_\text{h}$}
\put(105,53){Run R$_\text{ABC}$}
\put(14,65){(a) }
\put(87,65){(b) }
\put(12,9){(c) }
\put(88,9){(d) }
\end{picture}
          \caption{\label{fig:diffE21_13141516} Real part of polarization normalised by the energy spectrum for: (a) run R$_\text{nh}^1$; (b) run R$_\text{nh}^2$; (c) run R$_\text{h}$; (d) run R$_\text{ABC}$.}
\end{figure}

The third quantity, plotted in
Fig.~\ref{fig:diffE21_13141516} for runs in set R, is the directional anisotropy of the real part of polarization $\Re Z(k)$.
Since it is proportional to  the difference between poloidal and toroidal energy, as explained in section \ref{sec:modal} (see also \cite{bib:CMG, bib:favier,bib:delache2014}), when rotation is strong enough for wavevectors close to the horizontal plane to hold much more energy than wavevectors close to $\boldsymbol{\Omega}$ (which is the case for runs in set R, see Fig.~\ref{fig:direct_anis_1316}), this quantity provides information on the structure of turbulence at the considered scale.
In fact, Fig.~\ref{fig:diffE21_13141516} shows that for all the runs the real part of the spherically averaged polarization is negative at small wavenumbers (close to $k_F$), and positive at larger wavenumbers, which indicates that at large scales the toroidal energy is greater than the poloidal one, while at small scales the opposite happens. This is related to the presence of large scale ``vortical'' structures and of small scale ``jetal'' structures.
Upon comparing Figs. \ref{fig:diffE21_13141516}(a) and (b), one sees that increasing rotation increases the normalised polarization anisotropy, which is largest in the equatorial plane and vanishes in the axial direction, with a monotonous dependence in between.

In runs R$_\text{nh}^2$ (non-helical high-rotation Euler-forced, Fig. \ref{fig:diffE21_13141516}(b)), the real part of polarization reaches a maximum before decreasing towards the smallest dissipative scales, although not  reaching isotropy at the largest resolved wavenumber $k_\text{max}$.
In comparison, the slower rotating case (run R$_\text{nh}^1$) presented in Fig.~\ref{fig:diffE21_13141516}(a) maintains moderate polarization anisotropy down to the smallest scales, a behaviour similar to that of the helical case (R$_\text{h}$) of Fig. \ref{fig:diffE21_13141516}(c), even though the presence of helicity clearly increases small scale polarization anisotropy.
Run R$_\text{ABC}$ (moderate rotation ABC-forced, Fig. \ref{fig:diffE21_13141516}(d)), shows a polarization anisotropy level similar to that of run R$_\text{nh}^2$.

Therefore, although the relative helicity and the Reynolds and Rossby numbers of the ABC-forced run R$_\text{ABC}$ (Fig.~\ref{fig:diffE21_13141516}(d)) are similar to those of run $R_\text{h}$ (Fig.~\ref{fig:diffE21_13141516}(c)), the polarization anisotropy of the former is much higher than that of the latter. The level of this anisotropy for run R$_\text{ABC}$ is rather comparable to that obtained with a stronger rotation in Euler-forced runs (Fig. \ref{fig:diffE21_13141516}(b)).\\

\begin{figure}[]
\unitlength 1mm
\begin{picture}(140,55)(0,0)
\put(0,0){\includegraphics[width=70\unitlength]{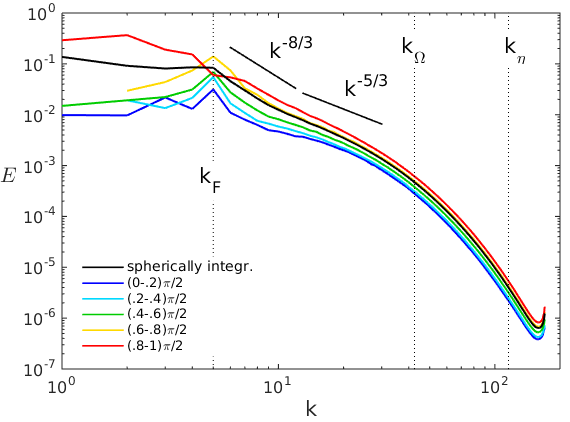}}
\put(75,0){ \includegraphics[width=70\unitlength]{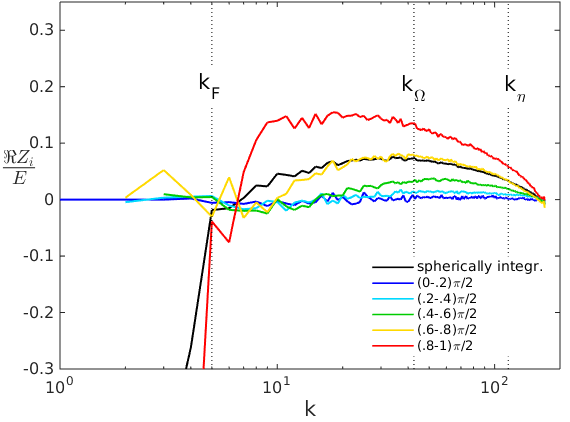}}
\put(35,9){(a)}
\put(110,9){(b)}
\end{picture}
          \caption{\label{fig:diffE21_r157} Helical shell-Euler forced run: the truncated system includes wavenumbers such that $4.6\le k\le 5.4$ and does not include the Coriolis force. The parameters of the simulation are $Re^\lambda=180$, $Ro^\omega=1.13$, $H_\text{rel}=0.50$. (a) Directional energy spectra; (b) real part of polarization normalised by the energy spectrum.}
\end{figure}
At this point, one may wonder if the differences observed between ABC-forced and Euler-forced simulations depend mainly on the intrinsic anisotropy of the ABC force or on the presence of an inverse cascade.
In order to definitely answer this question, we also performed a helical shell-Euler-forced rotating run (Fig.~\ref{fig:diffE21_r157}), in which the truncated system includes only modes corresponding to wavenumbers $k$ such that $4.6\le k\le 5.4$. 
Figure~\ref{fig:diffE21_r157}(a) shows the corresponding energy directional spectra and spherically-integrated spectrum, the slope of the latter is clearly stronger than $-5/3$.
Therefore, while an inverse cascade does not affect substantially energy directional anisotropy, the reason for a steeper energy spectrum slope in ABC forced runs is just the presence of an inverse cascade, absent in standard Euler-forced runs (forced for $k<k_F$). 
The polarization anisotropy for the shell-Euler-forced run, shown in Fig.~\ref{fig:diffE21_r157}(b), is stronger than that of the equivalent Euler-forced run without inverse cascade (Fig.~\ref{fig:diffE21_13141516}(c)), but smaller than the ABC-forced run (Fig.~\ref{fig:diffE21_13141516}(d)).
The increased polarization anisotropy in ABC-forced rotating runs therefore seems to be induced both by the intrinsic anisotropy of the ABC forcing and by the presence of an inverse cascade.

\subsection{Effect of rotation in higher Reynolds number cases}

In this section we study the anisotropy that naturally arises in the presence of background rotation through $1024^3$ resolution simulations, \textit{i.e.} considering runs in set S that have larger Reynolds numbers than those of set R, see Tab. \ref{tab:param2}. 
Run S$_\text{nh}^1$ also has smaller Zeman wavenumber, and thus permits to study the anisotropic features of scales much smaller than the Zeman scale.

\begin{figure}
        \centering
\unitlength 1mm
                \includegraphics[width=70\unitlength]{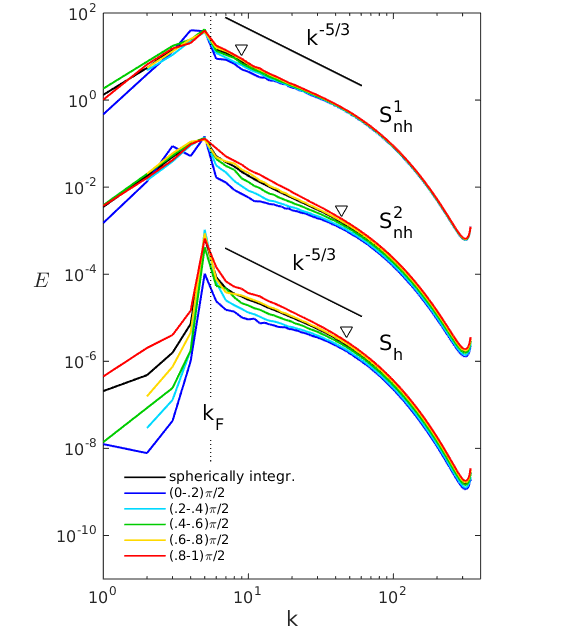}
         \caption{\label{fig:spettrm17} Directional energy spectra $E_i(k)$ for runs in set S. Markers indicate the Zeman scale \mbox{$k_\Omega=\sqrt{\left(2\Omega\right)^3/\epsilon}$}.}
\end{figure}

The directional energy spectra for runs of set S are plotted in Fig. \ref{fig:spettrm17}. A wide inertial range is observed, with a slope close to $-5/3$ for runs S$_\text{nh}^1$ and S$_\text{h}$ due to weak rotation.
At first glance, in the lowest rotation case (run S$_\text{nh}^1$) directional spectra collapse on the spherically integrated spectrum and small scales seem to recover isotropy.

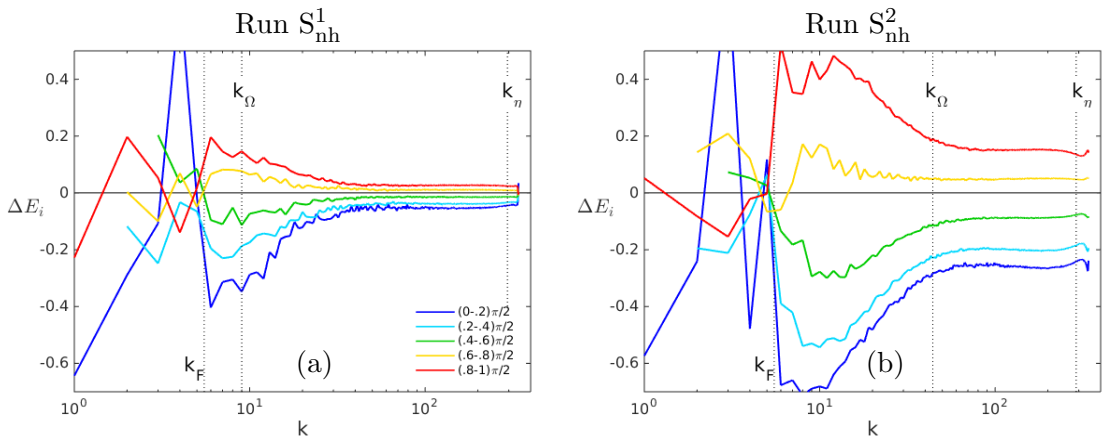
\begin{figure}[]
\unitlength 1mm
\begin{picture}(140,55)(0,0)
\put(0,0){\includegraphics[width=70\unitlength]{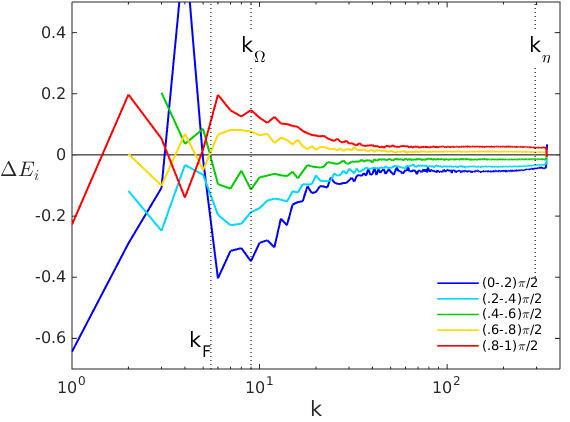}}
\put(75,0){\includegraphics[width=70\unitlength]{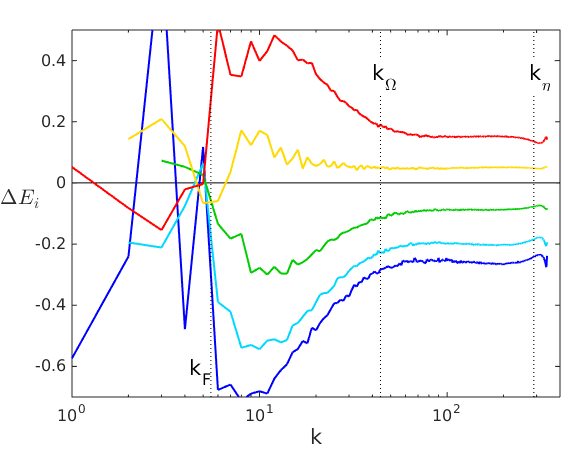}}
\put(38,9){(a)}
\put(113,9){(b)}
\put(30,54){Run S$_\text{nh}^1$}
\put(105,54){Run S$_\text{nh}^2$}
\end{picture}
         \caption{\label{fig:direct_anis_17} Energy directional anisotropy $\Delta E_i(k)$ for: (a) run S$_\text{nh}^1$; (b) run S$_\text{nh}^2$.}
\end{figure}

In Fig.~\ref{fig:direct_anis_17} we present the relative directional energy anisotropy for runs S$_\text{nh}^1$ and S$_\text{nh}^2$ (there is no substantial difference in energy directional anisotropy between runs S$_\text{h}$ and S$_\text{nh}^1$). 
Surprisingly, it shows that, notwithstanding the Reynolds number increase with respect to set R, the relative anisotropy stays roughly constant down to the smallest scales, after decreasing over the upper inertial spectral subrange. 
Even in the largest Rossby number case, run S$_\text{nh}^1$ (Fig. \ref{fig:direct_anis_17}(a)), the amplitude of the relative energy departure at small scales  is still significant and much larger than the anisotropy induced by forcing in absence of rotation (compare Fig. \ref{fig:direct_anis_17}(a) with Fig. \ref{fig:direct_anis_1234}(a)). 
A second important observation is that there seems to be two subranges in the inertial spectral range over which anisotropy behaves differently. In the first one (smallest wavenumbers), the relative anisotropy for all sectors decreases with wavenumber. Then, for wavenumbers greater than an intermediate value, the relative anisotropy remains roughly constant.
The separating wavenumber is clearly larger than $k_\Omega$ for run $S^1_\text{nh}$ (large Rossby number) and is close to $k_\Omega$ for the other runs, $S_\text{nh}^2$ and $S_\text{h}$, which have moderate Rossby numbers. 
Therefore, it is not clear how the separating scale between these two anisotropic ranges depends on the Zeman wavenumber.

\begin{figure}[]
\unitlength 1mm
\begin{picture}(140,105)(0,0)
\put(0,52){\includegraphics[width=70\unitlength]{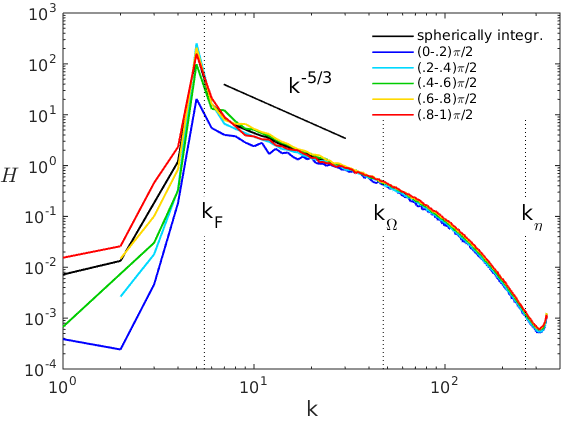}}
\put(75,52){\includegraphics[width=70\unitlength]{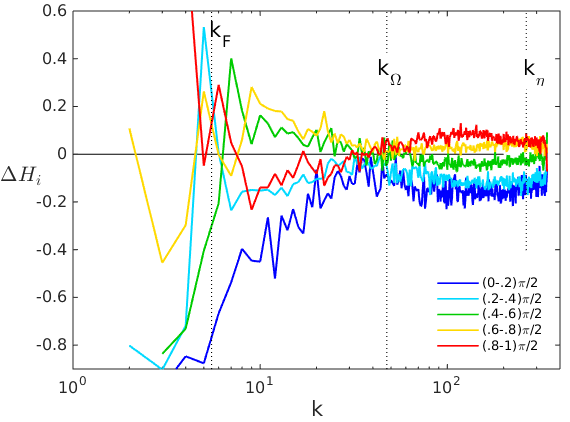}}
\put(35,0){\includegraphics[width=70\unitlength]{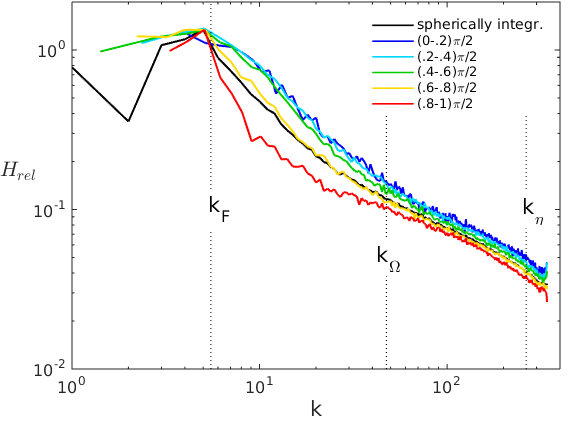}}
\put(35,61){(a)}
\put(110,61){(b)}
\put(70,9){(c)}
\end{picture}
         \caption{\label{fig:Hrel_r1002} (a) Directional helicity spectra $H(k,\theta)$, (b) helicity directional anisotropy $\Delta H_i(k)$, (c) relative helicity directional spectra ${H_\text{rel}}_i(k)=H_i(k)/\left( 2kE_i(k) \right)$, for run S$_\text{h}$.}
\end{figure}
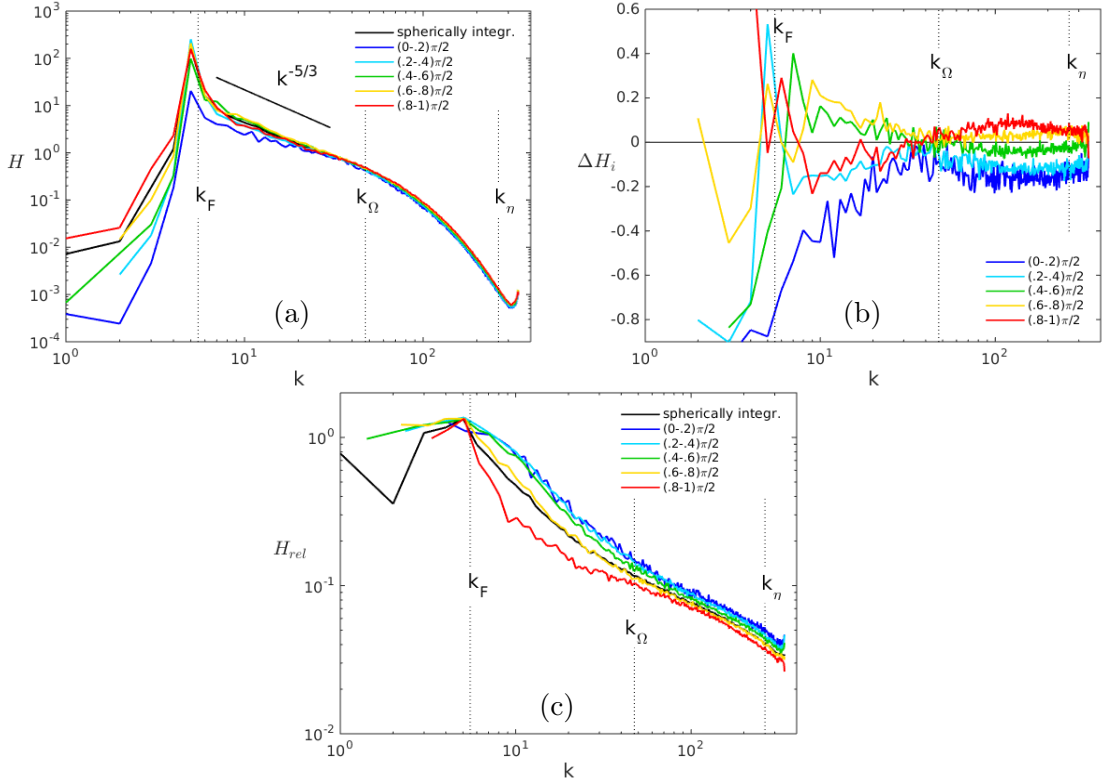

Finally, we present helicity directional spectra and directional anisotropy in Figs.~\ref{fig:Hrel_r1002}(a) and (b), respectively.
These figures show that helicity directional isotropy is reached at some intermediate wavenumber, but that it disappears for larger wavenumbers.
Similarly to energy, at small scales, sectors closer to the horizontal plane hold more helicity.
Figure~\ref{fig:Hrel_r1002}(c) also displays the relative helicity spectrum of every sector, ${H_\text{rel}}_i(k)=H_i(k)/\left( 2kE_i(k) \right)$.
From the viewpoint of relative helicity, no directional isotropy is recovered, and---no matter the scale---sectors closer to the horizontal plane hold lower relative helicity.
Therefore, even if both energy and helicity are concentrated in more horizontal wavevectors, the relative contents of helicity is larger for the less energetic and more vertical wavevectors.

\section{Threshold wavenumber between two anisotropic ranges}
\label{sec:kT}
Recall from the introduction that if the kinematic viscosity $\nu$ tends to zero (and the Reynolds number tends to infinity), both $\eta$ and $r_{\Omega d}$ tend to zero. The only relevant small-scale characteristic lengthscale is then the Zeman scale $r_\Omega$, which is the scale at which the characteristic rotation time equals the characteristic inertia time.
For this reason, according to classical dimensional arguments \cite{bib:mininni2012,bib:delache2014,bib:zeman}, in the asymptotically inviscid limit, scales much larger than $r_\Omega$ should be strongly affected by rotation and should therefore be more anisotropic, while scales much smaller than $r_\Omega$ are expected to be dominated by the nonlinear dynamics and to have isotropic properties.
However, only finite Reynolds number turbulence can be tackled through simulations and experiments, and very large Reynolds numbers are needed to achieve a good scale separation.
DNS by \cite{bib:delache2014, bib:mininni2012} seem to confirm isotropy recovery at small scales, while in experiments by \cite{bib:lamriben2011} the anisotropy is found to be stronger at small scales.
In particular, in the forced rotating simulation of \cite{bib:mininni2012} isotropization seems to occur at a precise wavenumber (close to $k_\Omega$). In \cite{bib:delache2014}, in which decaying rotating turbulence is investigated, isotropy is recovered only if rotation is weak enough, and a link between $k_\Omega$ and the wavenumber corresponding to maximum anisotropy is observed.
Therefore, both the anisotropic character of small scales and the role of the Zeman scale are not fully understood.

In section \ref{sec:results2}, our analysis---that uses normalised indicators and includes simulations with different Rossby numbers---shows no isotropy recovery, in contrast with previous numerical results \cite{bib:delache2014, bib:mininni2012} but in agreement with experiments \cite{bib:lamriben2011}.
Nevertheless, even if isotropy is not recovered at small scales in our simulations, two different anisotropic ranges with qualitatively different anisotropic features can be identified, see \textit{e.g.} Figs.~\ref{fig:spettrm17} and \ref{fig:direct_anis_17}.
The low-wavenumber range shows large anisotropy decreasing with wavenumber, while the anisotropy level at larger wavenumbers is significantly lower, although not zero.
Then, one may wonder if the threshold wavenumber between these two ranges has a specific physical interpretation.
In order to answer this question, we analyse here a larger number of Euler-forced runs (17 runs with $512^3$ resolution and 6 runs with $1024^3$ resolution), with $Ro^\omega$ ranging from $0.69$ to $9.6$, $Re^\lambda$ ranging from $73.9$ to $414$, and scale separation $r_\Omega/\eta$ ranging from $1.3$ to $68$.
Note that this set also includes runs with different forcing scales ($k_F=1.5$, $3.5$ and $5.5$), 
different relative helicity (ranging from $0$ to $0.84$), and runs that include or do not include the Coriolis force in the spherically truncated system.

First, we define a systematic method to compute the threshold wavenumber $k_T$, separating \sloppy{small-wavenumber} (large anisotropy) and large-wavenumber (low anisotropy) ranges.
Then, we investigate its dependence on the other parameters of the flow and look for a physical interpretation for $k_T$.

\begin{figure}[h]
\centering
\unitlength 1mm
\includegraphics[width=70\unitlength]{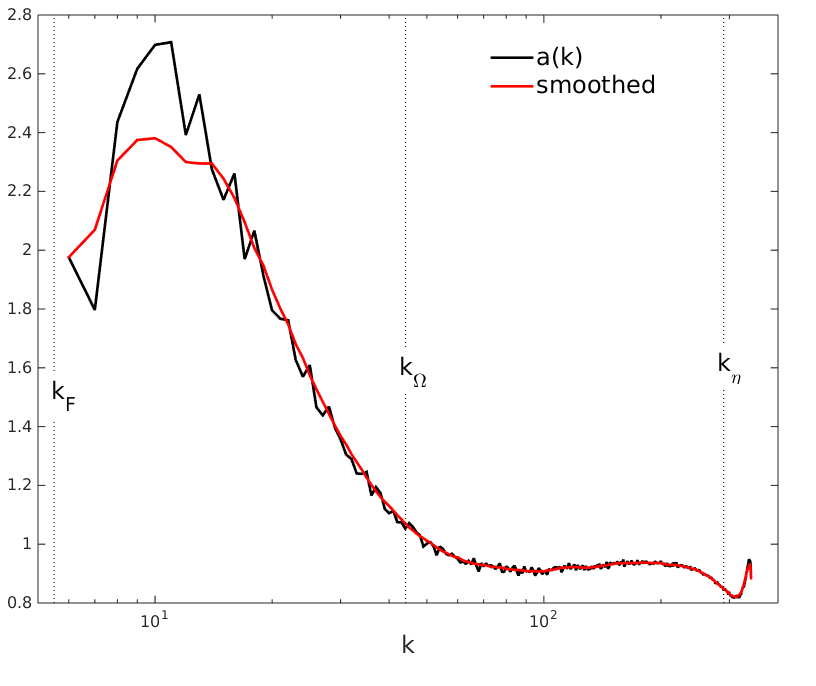}
\caption{\label{fig:a_k}Anisotropy indicator (defined by Eq. \eqref{eq:a_k}) for run S$_\text{nh}^2$.}
\end{figure}

Since for every run five energy directional-anisotropy indicators $\Delta E_i(k)$ are available, we first reduce them to a single indicator $a(k)$. In particular, we normalise every $\Delta E_i(k)$ by its mean value over the range $k>k_F$, and then average them:
\begin{equation}
a(k)=\frac{1}{5}\sum\limits_{i=1}^5 \frac{\Delta E_i(k)}{\left<\Delta E_i\right>}.
\label{eq:a_k}
\end{equation}
Figure~\ref{fig:a_k} shows the anisotropy indicator $a(k)$ corresponding to run S$_\text{nh}^2$ (Fig.~\ref{fig:direct_anis_17}(b)).
In all rotating runs we found that $a(k)$ quickly decreases with wavenumber at large scales, reaches a minimum and then slowly increases with wavenumber up to the dissipative scales. Therefore, we compute $k_T$ as the wavenumber corresponding to the minimum of $a(k)$, after possible smoothing.

\begin{figure}[]
\unitlength 1mm
\begin{picture}(140,63)(0,0)
\put(0,0){\includegraphics[width=66\unitlength]{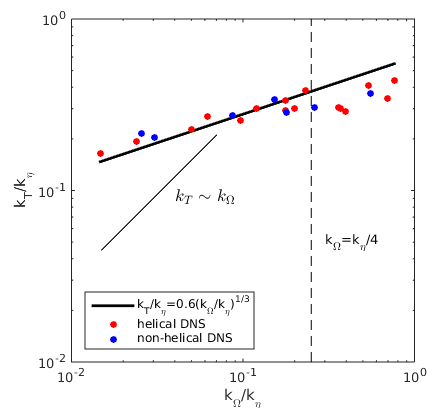}}
\put(70,0){\includegraphics[width=66\unitlength]{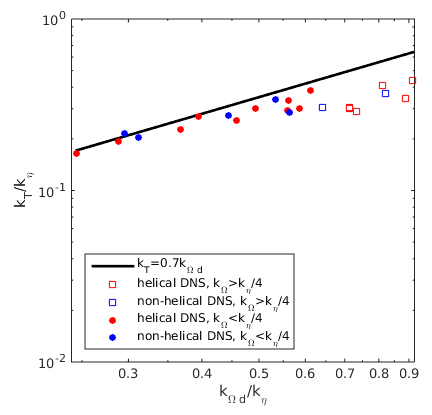}}
\put(48,9){(a)}
\put(118,9){(b)}
\end{picture}
        \caption{\label{fig:kTketa_komdketa}$k_T/k_\eta$ plotted as a function of (a) $k_\Omega/k_\eta$, (b) $k_{\Omega d}/k_\eta$. For comparison, the slope corresponding to $k_T\sim k_\Omega$ is shown too.}
\end{figure}

As a first attempt, it is natural to investigate the dependence of $k_T$ on the Zeman wavenumber $k_\Omega$, with the purpose of checking the existence of a range in which $k_T\sim k_\Omega$.
In Fig.~\ref{fig:kTketa_komdketa}(a), $k_T/k_\eta$ is plotted as a function of $k_\Omega/k_\eta$. For $k_\Omega/k_\eta\lesssim 1/4$ (weak or moderate rotation), $k_T/k_\eta$ clearly increases with $k_\Omega/k_\eta$, with a power law of exponent $1/3$. For larger values of $k_\Omega/k_\eta$, markers are more scattered, and no clear trend is observed. 
One possible explanation for the existence of these two regimes is that, if rotation is too strong (or equivalently $k_\Omega/k_\eta$ is too large), the threshold wavenumber $k_T$ is located in the dissipative range, whereas in the opposite case it is in the inertial range. These two ranges are phenomenologically different, and different laws can be expected in the two cases. The rest of our discussion will be performed in the regime $k_\Omega/k_\eta\lesssim 1/4$, in which $k_T/k_\eta\sim(k_\Omega/k_\eta)^{1/3}$. This amounts to discard the lowest Rossby number runs.

In brief, Fig.~\ref{fig:kTketa_komdketa}(a) shows two important results: first, depending on the closeness of $k_\Omega$  to $k_\eta$ two subranges with different behaviours are observed and second, in the low $k_\Omega$ range, $k_T$ scales as $k_\Omega^{1/3}k_\eta^{2/3}$. In this regime, $k_T$ is therefore not proportional to $k_\Omega$, and depends on the dissipative scale as well.
Recalling from the introduction that, from the definitions of $k_{\Omega d}$, $k_\Omega$ and $k_\eta$, $k_{\Omega d}=k_\Omega^{1/3}k_\eta^{2/3}$, this means that $k_T$ scales as $k_{\Omega d}$. This result is confirmed by Fig.~\ref{fig:kTketa_komdketa}(b), which furthermore shows that the factor between $k_T$ and $k_{\Omega d}$ is close to 1, therefore:
\begin{equation}
k_T\approx k_{\Omega d}=\left( \frac{2\Omega}{\nu} \right)^{1/2}.
\label{eq:scalawkT}
\end{equation}
\begin{figure}[]
\unitlength 1mm
\begin{picture}(140,63)(0,0)
\put(0,0){\includegraphics[width=66\unitlength]{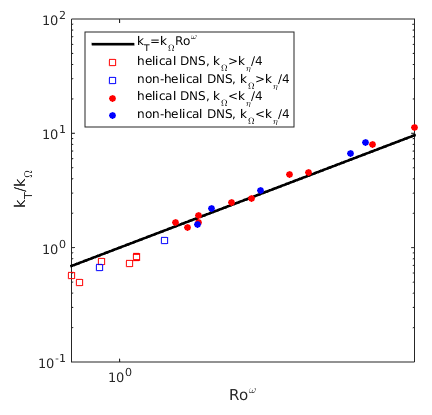}}
\put(70,0){\includegraphics[width=66\unitlength]{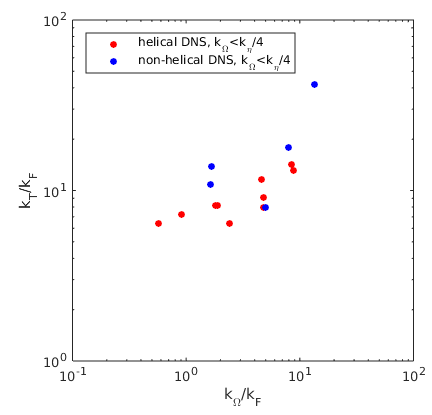}}
\put(35,9){(a)}
\put(105,9){(b)}
\end{picture}
        \caption{\label{fig:kTkF_komkF}(a) $k_T/k_\Omega$ plotted as a function of $Ro^\omega$, (b) $k_T/k_F$ plotted as a funcion of $k_\Omega/k_F$.}
\end{figure}
This relation definitely identifies $k_T$ as the wavenumber at which the rotation time equals the characteristic dissipation time, provided that $k_\Omega/k_\eta$ is not too large (in practice, $k_\Omega \lesssim k_\eta/4$).
In other words, at small wavenumbers anisotropy quickly decreases with the wavenumber, then reaches a minimum at $k\approx k_{\Omega d}$, after which it slowly increases up to the dissipative scales.
Also recalling from the introduction that, under the hypothesis $\omega'\sim \nu k_\eta^2$, $Ro^\omega$ should scale as $(k_\eta/k_\Omega)^{2/3}$, the above result, Eq.~\eqref{eq:scalawkT} yields: $k_T\sim k_\Omega Ro^\omega$.
To check this, $k_T/k_\Omega$ is plotted as a function of $Ro^\omega$ in Fig.~\ref{fig:kTkF_komkF}(a). Again, this scaling is satisfied for the data corresponding to $k_\Omega< k_\eta/4$, and the proportionality factor is close to 1.

In order to make sure that the scaling law found above is not artificially induced by the forcing, further investigation is required.
In fact, if rotation is too weak, the threshold wavenumber $k_T$ may be close enough to $k_F$ for the forcing scheme to affect its value.
In Fig.~\ref{fig:kTkF_komkF}(b), $k_T/k_F$ is plotted as a function of $k_\Omega/k_F$ (runs for which $k_\Omega> k_\eta/4$ are not included). No trend is visible from these data, so that no forcing effect is detected. Such an effect might, however, be evidenced in simulations with larger forcing wavenumber or larger Rossby number.

\begin{figure}[!t]
\centering
\unitlength 1mm
    \includegraphics[width=66\unitlength]{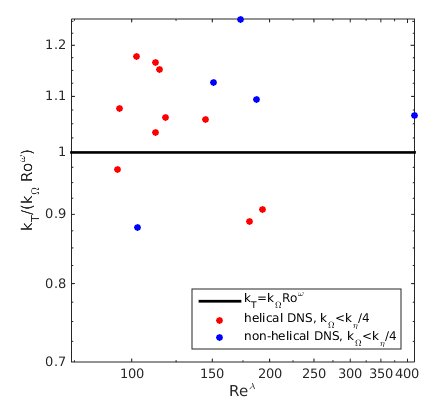}
         \caption{\label{fig:dipendRelam}Dependence of the $k_T$ scaling law on the Reynolds number.}
\end{figure}

Finally, we investigate the dependence of the $k_T$ scaling law on the Reynolds number, see Fig.~\ref{fig:dipendRelam} in which $k_T/\left(k_\Omega Ro^\omega\right)$ is plotted as a function of $Re^\lambda$.
As already shown in Fig. \ref{fig:kTkF_komkF}(a), this quantity is always close to one. Moreover, there is no correlation between it and $Re^\lambda$.
It seems therefore that, in the range covered by our runs, the scaling law of $k_T$ (Eq. \eqref{eq:scalawkT}) does not depend on the Reynolds number.

Note that in the asymptotic inviscid limit, according to our scaling law, \mbox{$k_T\sim k_{\Omega d} \rightarrow \infty$} and thus only the low-wavenumber anisotropic range ($k<k_T$) should persist.
In this range anisotropy decreases with wavenumber, which is possibly consistent with the classical argument according to which isotropy should be recovered at scales infinitely smaller than the Zeman scale (if the minimum of $a(k)$ tends to zero). 

We did not compare the results presented in this section to existing results in the literature \cite{bib:delache2014, bib:mininni2012,bib:lamriben2011} because of lack of the required data in these articles.

\section{Summary and conclusions}\label{sec:concl}
In this work, we have investigated the effect of three large-scale spectral forcing methods, namely the Euler, the ABC and the negative viscosity forcing schemes, on the anisotropy of turbulence. We have first considered the case of turbulence believed to be isotropic and we have quantified the scale dependent anisotropy of the flow, before considering rotating turbulence in which anisotropy is naturally produced by the action of the Coriolis force.

Since isotropy or anisotropy of turbulence concerns all the scales in the flow, we have proposed to not merely quantify it by one-point statistics, but have instead considered multiscale statistics. We have thus considered refined two-point statistics by decomposing the spectral velocity tensor into different contributions: energy, helicity and polarization spectral densities. 
Moreover, we have computed directional spectra by partial integration of these spectral densities over five sectors.
The directional energy spectra allow for instance to distinguish trends towards bidimensionalization or vertically-sheared horizontal flows. Helicity spectra further indicate the helical contents at the considered scale or wavenumber. And finally, the less commonly used polarization spectrum testifies of the local structure of the flow at a given scale, through the difference between toroidal and poloidal energy. Overall, these three direction-dependent spectra contain the complete information for characterizing axisymmetric turbulence at the level of two-point velocity correlations.

First, for non-rotating turbulence, we have shown that energy and helicity directional anisotropies can arise at all scales under the effect of forcing, when the number of excited modes is too low or when few of the forcing modes hold much more energy than the others. As a consequence, the ABC forcing scheme always affects  directional anisotropy since, regardless of the forcing wavenumber $k_F$, it only excites six modes.
On the contrary, the Euler forcing is not bound to exciting a limited number of modes. In fact, even if the relative helicity is very large (so that the energy is concentrated in the largest forced wavenumbers), one can always set a suitable value of $k_F$ such that the induced anisotropy is negligible.
Our implementation of Euler forcing is original in the sense that it can be achieved with any choice of $k_F$ and the amount of helicity injected in the flow can be controlled, whereas previous implementations were limited to $k_F=1.5$ and non-helical turbulence.
We have shown that polarization directional anisotropy can develop as well in forced turbulence expected to be isotropic, but that it gradually decreases at increasing wavenumber so that it is negligible at small scales.

Second, we considered forced homogeneous turbulence subject to external rotation. The flow dynamics is then influenced both by the large-scale synthetic forcing and by the background rotation.
We showed that in the ABC-forced rotating simulations the energy spectrum slope is altered.
This is due to the fact that, in presence of rotation, the ABC forcing allows energy to cascade backward.
Furthermore, the polarization anisotropy level is similar to that obtained at lower Rossby numbers in Euler-forced runs with no inverse cascade.
This last result is partly due to the anisotropic nature of the ABC forcing and partly to the inverse cascade.

We then showed that in rotating turbulence, energy and helicity directional anisotropies are present at the smallest scales of the flow even at large Rossby numbers (even though the anisotropy level decreases at increasing Rossby number).
However, two different wavenumber ranges, in which anisotropy evolves differently, were evidenced: directional anisotropy decreases at increasing wavenumber at large scales, then becomes minimal at an intermediate wavenumber before slowly increasing with wavenumber up to the dissipative scales.
The characteristic lengthscale separating these two ranges is not the Zeman scale (at which rotation effects are of the order as inertial ones). When it is large enough, we rather identified it as the scale at which dissipative effects are of the same order as those of rotation.
This provides not only a qualitative but also an accurate quantitative threshold separating the two anisotropic subranges.
This behaviour is observed consistently at all the Reynolds numbers and for all the different configurations we have examined.
In the asymptotic limit of infinite Reynolds number, our results predict anisotropy to monotonically decrease at increasing wavenumber, a scenario possibly consistent with the classical dimensional argument according to which isotropy should be recovered at scales infinitely smaller than the Zeman scale.

\section*{Acknowledgements}
We are grateful to Prof. Y. Kaneda and his team for providing the $2048^3$ data field of isotropic turbulence forced by the negative viscosity method. The other data were computed using computational time provided by GENCI/CINES under project number A0012A2206, and by the FLMSN HPC mesocenter in Lyon, France (PSMN [\'Ecole Normale Sup\'erieure de Lyon], P2CHPD [Universit\'e Claude Bernard Lyon 1] and PMCS2I [\'Ecole Centrale de Lyon]), funded by the equip@Meso project.

\bibliographystyle{abbrv}
\bibliography{refs1.bib}

\end{document}